\documentclass[journal,10pt]{IEEEtran}
\usepackage{algorithm,algpseudocode,graphicx}
\usepackage{tikz}
\usetikzlibrary{shapes,arrows}
\usepackage{pstricks}
\usepackage{pst-node,pst-blur}
\usetikzlibrary{arrows}
\usepackage{amsfonts}
\usepackage{tabularx}
\usepackage{subcaption}
\usepackage{amssymb}
\usepackage{booktabs}
\usepackage{balance}
\usepackage[figurename=Fig., labelsep=period, font=small]{caption}
\usepackage{multirow}
\usepackage{flushend}
\usepackage{epsfig}
\usepackage{epstopdf}
\usepackage{arydshln}
\usepackage{tikz,pgfplots}
\usepgfplotslibrary{groupplots}
\usetikzlibrary{arrows, arrows.meta, backgrounds, calc, intersections, decorations.markings, decorations.pathreplacing, positioning, shapes}
\pgfplotsset{compat=newest}

\newcommand{\bi}{\begin{itemize}}
\newcommand{\ei}{\end{itemize}}

\newcommand{\be}{\begin{enumerate}}
\newcommand{\ee}{\end{enumerate}}
\newcommand{\bd}{\begin{description}}
\newcommand{\ed}{\end{description}}
\newcommand{\bc}{\begin{center}}
\newcommand{\ec}{\end{center}}
\newcommand{\bt}{\begin{tabbing}}
\newcommand{\et}{\end{tabbing}}
\newcommand{\bfig}{\begin{figure}}
\newcommand{\efig}{\end{figure}}
\newcommand{\beq}{\begin{equation}}
\newcommand{\beqarr}{\begin{eqnarray}}
\newcommand{\beqarrn}{\begin{eqnarray*}}
\newcommand{\eeq}{\end{equation}}
\newcommand{\eeqarr}{\end{eqnarray}}
\newcommand{\eeqarrn}{\end{eqnarray*}}
\newcommand{\bflr}{\begin{flushright}\vspace{-0.2in}}
\newcommand{\eflr}{\end{flushright}}
\newcommand{\bsub}{\begin{subequations}}
\newcommand{\esub}{\end{subequations}}
\newcommand{\barr}{\begin{array}}
\newcommand{\earr}{\end{array}}
\newcommand{\nn}{\nonumber}
\def\binom#1#2{\left( \!  \barr{c} #1 \\ #2 \earr \!  \right)}

\def\undb#1{\mbox{\bf{#1}}}

\def\dn{\stackrel{\scriptscriptstyle \triangle}{=}}

\usepackage[noadjust]{cite}

{\newtheorem{proposition}{Proposition}}

\usepackage[cmex10]{amsmath}

\allowdisplaybreaks  
   
 \makeatletter 
 \def\@eqnnum{{\normalsize \normalcolor (\theequation)}} 
  \makeatother
\begin{document}
\title{SEP Analysis of Fluid Antenna GSC Reception under Correlated Nakagami-$m$ Fading and Imperfect Channel Estimation}
\author{Soumya P. Dash,~\IEEEmembership{Senior~Member,~IEEE}, Italo Atzeni,~\IEEEmembership{Senior~Member,~IEEE}, \\ and George C. Alexandropoulos,~\IEEEmembership{Senior~Member,~IEEE}
\thanks{A preliminary version of this paper has been presented at the IEEE SPAWC 2026, Athens, Greece~\cite{dash2026symbolerroranalysisfluid}.}
\thanks{S. P. Dash is with the School of Electrical and Computer Sciences, Indian Institute of Technology Bhubaneswar, Argul, Khordha, Odisha, 752051 India, and with the Centre for Wireless Communications, University of Oulu, 90570 Oulu, Finland (e-mail: soumyapdashiitbbs@gmail.com).}
\thanks{I. Atzeni is with the Centre for Wireless Communications, University of Oulu, 90570 Oulu, Finland (e-mail: italo.atzeni@oulu.fi).}
\thanks{G. C. Alexandropoulos is with the Department of Informatics and Telecommunications, National and Kapodistrian University of Athens, Panepistimiopolis Ilissia, 16122 Athens, Greece, and also with the Department of Electrical and Computer Engineering, University of Illinois Chicago, IL 60601, USA (e-mail: alexandg@di.uoa.gr).}
}
\maketitle
\newcounter{my_counter}

\begin{abstract}
Channel estimation (CE) errors and correlation among the fading channels are two of the most fundamental factors degrading the performance of any multi-antenna system. This paper aims to analytically capture their joint effect on fluid antenna systems (FASs) performing generalized selection combining (GSC), i.e., combining the received signals at the best $\tilde{L}$ of the $L$ available ports using maximum ratio combining (MRC). More specifically, the FAS receiver is subject to arbitrarily correlated Nakagami-$m$ fading channels following the fully correlated Jakes' model, and employs a pilot-aided maximum-likelihood rule for CE prior to data detection. A novel tight moment-matching-based Gamma approximation is proposed for the marginal distribution of the estimated channel gains, which is then used to derive the joint statistics of the estimated correlated channels for the case of non-identical integer or half-integer fading severity parameters. Considering that the transmitter employs $M$-ary phase-shift keying or quadrature amplitude modulation, the joint statistics are further exploited to obtain exact finite-integral symbol error probability (SEP) expressions for a GSC-based FAS receiver, along with closed-form counterparts for the case of identical fading parameters. In addition, asymptotic SEP expressions at high signal-to-noise ratio are derived, analytically establishing that non-ideal CE and fading correlation degrade only the FAS's coding gain, while the diversity order is preserved and governed solely by the number of combined ports and the estimated fading severity. The presented numerical SEP investigations offer concrete design insights into the trade-offs between CE quality, fading correlation, and port selection for the GSC-based FAS receiver.
\end{abstract}
\IEEEpeerreviewmaketitle
 
\section{Introduction}
\IEEEPARstart{T}{he} sixth generation (6G) of wireless networks is envisioned to deliver unprecedented spectral efficiency, reliability, and connectivity density~\cite{saad2020vision}. Multiple-input multiple-output (MIMO) technology, and massive MIMO in particular, has been central to this progress by exploiting large numbers of antenna elements~\cite{marzetta2016massive}. However, the size, power, and cost of the radio-frequency (RF) chains required by conventional arrays in traditional antenna systems limit their use in compact devices. In this context, fluid antenna systems (FASs) have emerged as a flexible alternative~\cite{new2024tutorial}. FASs dynamically switch a single or a few RF chains among many densely packed radiating positions, called \emph{ports}, within a small aperture, thereby harvesting the spatial diversity of the environment with a much smaller hardware footprint. Thus, FAS requires an accurate and tractable performance characterization under practical operating conditions for their deployment~\cite{11455595, ganesh2025OPMRC}.

The performance of FASs has been studied extensively under perfect channel state information (CSI), with a focus on proposing approximations to Jakes' model describing the correlation among their ports. The closed-form outage probability and diversity gain were derived in~\cite{new2024insights}, and ergodic capacity and multiple-access extensions were reported in~\cite{wong2021performance, wong2022fama}. The diversity and coded-modulation design of FAS was investigated in~\cite{psomas2024diversity}. A central difficulty is that closely spaced ports are strongly correlated. Accordingly, closed-form spatial-correlation parameters~\cite{wong2022closedform}, tractable approximations of the FAS channel~\cite{khammassi2023new}, and block-correlation models~\cite{ramirez2024spatial} have been proposed. Beyond Rayleigh fading, analyses have considered correlated Nakagami-$m$ in~\cite{tlebaldiyeva2022enhancing, vegasanchez2024simple} and $\alpha$-$\mu$ fading in~\cite{alvim2024alphamu}. However, all of these works assume that the FAS receiver knows the channel gain at every port exactly, so that port selection and combining are error-free.

Realizing the gains promised by FASs in practice, however, hinges on accurate CSI at the receiver, which must itself be acquired. Channel estimation (CE) is typically performed via pilot transmissions, which introduces estimation errors that degrade both the combining weights computation and the subsequent detection process~\cite{biguesh2006training}. Moreover, the training overhead grows with the number of ports. This has motivated a growing line of work on CE for FASs. Linear minimum mean squared error (MMSE) estimation was combined with port skipping and port selection to analyze the outage performance of large-scale FASs in~\cite{skouroumounis2022largescale, skouroumounis2023lmmse}. Prior-aided Bayesian regression over the spatial channel process was used in~\cite{zhang2024sbar}, and sparse-recovery and maximum-likelihood (ML) estimators that exploit the spatial correlation of ports were developed in~\cite{chen2025isamv}. The sampling distance and training cost needed to reconstruct the full port channel were studied in~\cite{xu2024oversampling}, whereas pilot-aided MMSE estimation with its error statistics was analyzed for large FASs in~\cite{ghadi2026efas}. These contributions focus on estimator design, reconstruction accuracy, and training overhead. Moreover, the performance has been mainly expressed in terms of either the outage probability or achievable rate, while the analysis of the symbol error probability (SEP) has been scarce. In particular, existing work on SEP for FASs is quite limited \cite{11360610, dash2026symbolerroranalysisfluid}, and assumes perfect CSI at the receiver for data detection. However, the FAS performance under correlated fading and pilot-aided estimation of the channel gains is rather crucial, especially when combining a subset of available ports based on imperfect channel estimates. This practical setting requires the joint statistics of the estimated correlated channel gains, and, in addition to degrading detection, the estimation error corrupts both the port selection and combining.

Outside the FAS literature, the impact of imperfect CSI on diversity combining over Nakagami-$m$ fading (widely adopted to model fading channels owing to its versatility~\cite{5411725,6184250,ganeshgc25}, as it encompasses the Rayleigh distribution as a special case and closely fits a broad range of empirically observed propagation conditions) has been studied for conventional maximum ratio combining (MRC) receivers. Examples include optimized combining with pilot-aided estimates~\cite{mallik2006optimized}, $M$-ary quadrature amplitude modulation ($M$-QAM) with MRC and equal-gain combining (EGC) and estimation errors~\cite{yao2007effect}, and bit error rate analysis with imperfect CSI~\cite{jacobs2008exact}. The effect of CE on subset diversity, where the best branches are selected and combined, was characterized exactly for independent and identically distributed Rayleigh fading in~\cite{gifford2008antenna}. Analyses of MRC and MRC-like (e.g., orthogonal space-time block codes (OSTBCs)) architectures subject to correlated Nakagami-$m$ fading and perfect CSI were presented in~\cite{lombardo1991mrc, win2000mrc,4786505,5682425, alexandropoulos2017swc}. For arbitrarily distributed and correlated channels, the bit error rate for the transmission of OSTBCs was analyzed in \cite{jacobs2011analysis}. However, for correlated Nakagami-$m$ branches, the joint effect of CE error and fading correlation was addressed only approximately in~\cite{alexandropoulos2010adaptive}, whereas the analysis in~\cite{jacobs2011analysis} requires computationally demanding four-fold integration and is limited to a restricted set of modulations. This status indicates a lack of tractable frameworks for performance analysis of diversity reception in the practical setting of arbitrarily correlated fading and imperfect CSI.

The above review on FAS performance analysis exposes a similar research gap. There exist exact analyses on perfect CSI, whereas the CE works treat estimation accuracy and overhead but not the resulting SEP of a generalized selection combining (GSC) diversity scheme. In turn, the classical diversity analyses with imperfect CSI also do not cover correlated, selection-based reception. Consequently, the impact of CE errors on the error performance of a practical FAS, whose strongly correlated ports are selected and combined using estimated gains, remains uncharacterized. To address this research gap, this paper focuses on the SEP performance analysis of a GSC-based FAS receiver operating under arbitrarily correlated Nakagami-$m$ fading, following Jakes' model for port correlation, and performing realistic pilot-aided CE. Specifically, the main contributions are summarized as follows:
\begin{itemize}
\item A novel tight moment-matching-based Gamma approximation for the marginal distribution of squared Nakagami-$m$ channel gains, resulting from ML estimation through pilot transmissions, is presented. The proposed approximate distribution is valid for arbitrary integer or half-integer (IoHI) values of the $m$-parameter denoting the fading severity.
\item A novel closed-form approximation is derived for the correlation coefficient between any pair of estimated channel gains at the FAS ports, showing that CE errors provably reduce the effective correlation relative to the underlying actual Nakagami-$m$ channels.
\item The derived joint distribution of estimated arbitrarily correlated Nakagami-$m$ random variables (RVs) is used to obtain novel exact SEP expressions, requiring only finite-range integration, for a GSC-based FAS receiver under both $M$-ary phase-shift keying (PSK) and $M$-QAM signaling constellations and imperfect CSI, together with closed-form expressions for the practically relevant case of identically distributed ports.
\item Asymptotic SEP expressions at high signal-to-noise ratio (SNR) are also presented, showing analytically that non-ideal CE and fading correlation degrade only the FAS coding gain, while the diversity order is preserved and depends solely on the number of combined ports and the fading severity of the estimated channels.
\item All derived expressions are validated through Monte Carlo simulations, and extensive numerical results quantify the joint impact of CE quality, fading severity, correlation, and port-selection ratio on the SEP performance of a GSC-based FAS receiver. Our results also include comparisons against conventional MRC reception.
\end{itemize}

The remainder of this paper is organized as follows. Section~II presents the considered FAS system and channel models as well as the adopted CE framework. Section~III derives the joint statistics of the estimated correlated Nakagami-$m$ port channels. Section~IV presents the SEP analysis, including asymptotic results. Section~V provides numerical and simulation results, while Section~VI concludes the paper.

\textit{Notations:} The $i$-th element of the vector $\mathbf{a}$ and the $(i,j)$-th element of matrix $\mathbf{A}$ are represented by $[\mathbf{a}]_i$ and $[\mathbf{A}]_{i,j}$, respectively, while $(\cdot)^{\mathrm{T}}$ denotes the transpose operator. $\sqrt{\mathbf{A}}$ is a matrix having as elements the square roots of the elements of $\mathbf{A}$. $x \sim \mathcal{N}(0,\sigma^2)$ or $x \sim \mathcal{CN}(0,\sigma^2)$ indicates that $x$ is a Gaussian or circularly symmetric complex Gaussian RV, respectively, with zero mean and variance $\sigma^2$. $\mathrm{Re}(z)$ and $\mathrm{Im}(z)$ represent the real and imaginary parts of the complex number $z$, while $|z|$ denotes its magnitude. $\mathbb{E}[\cdot]$ represents expectation,  $\Gamma(\cdot)$ is the Gamma function~\cite[eq.~(8.310/1)]{gradshteyn2000table}, and $B(x,y)$ is the Beta function. $\mathbb{R}$ denotes the set of real numbers, and $\jmath \triangleq \sqrt{-1}$ is the imaginary unit.

\section{System and Channel Models}
\label{Sec:Model}
We consider a multi-antenna receiver communicating with a single-antenna transmitter over a slow-varying, frequency non-selective wireless environment. The receiver is equipped with a one-dimensional FAS with $L$ evenly spaced ports over a linear length of $W \lambda$, where $\lambda$ is the signal's wavelength. The received signal at the $\ell$-th port can be mathematically expressed as
\beq
r_\ell \triangleq h_\ell s + n_\ell, \quad \forall \ell=1,2,\ldots,L ,
\label{eq:rx_FAS}
\eeq
where $s$ is the transmitted complex-valued data symbol chosen from a discrete modulation set and having average energy $E_{\text{s}}$, $h_\ell$ is the $\ell$-th complex-valued element of the channel gain vector $\mathbf{h}\triangleq \left[h_1,h_2,\ldots,h_L \right]^{\mathrm{T}}$, and $n_{\ell} \sim \mathcal{CN}(0,\sigma_{\text{n}}^2)$ is the additive white Gaussian noise (AWGN) at the $\ell$-th port. We make the usual assumption that $\{n_\ell\}$ are statistically independent over all ports and with respect to $s$.
\subsection{Channel Model}
Let $\gamma_\ell \triangleq \left|h_\ell\right|^2E_{\text{s}}/\sigma_{\text{n}}^2$ be the instantaneous received SNR per symbol at the $\ell$-th port of the FAS receiver. Moreover, we assume that the channel from the transmitter to each port of the FAS follows the Nakagami-$m$ distribution~\cite{yacoub2005nakagami}, i.e.,  $\big\{\left|h_\ell\right|\big\}$ are Nakagami-$m$-distributed RVs, each with fading parameter $m_\ell \geq 0.5$. The marginal probability density function (PDF) of each $\gamma_\ell$ is given by
\beq
f_{\gamma_\ell} \left( x \right) \triangleq \frac{m_\ell^{m_\ell} x^{m_\ell-1}}
{\overline{\gamma}^{m_\ell} \Gamma \left(m_\ell\right)}
\exp \left(- \frac{m_\ell x}{\overline{\gamma}} \right), \quad x \geq 0,
\label{eq:pdf_TAS}
\eeq
where $\overline{\gamma} \triangleq \mathbb{E}[|h_\ell|^2]E_{\text{s}}/\sigma_{\text{n}}^2$ represents the common actual average received SNR per FAS port.

In this paper, we adopt Jakes' model for the correlation coefficient between the channel gains of any two ports of the FAS receiver~\cite{wong2021fluid}. To this end, the correlation matrix (CM) of $\mathbf{h}$, denoted by $\mathbf{\Sigma} \in\mathbb{R}^{L\times L}_+$, is symmetric positive definite and is expressed as
\begin{equation}\label{eq:CM_FAS}
\left[\mathbf{\Sigma}\right]_{k,\ell} \triangleq
\left\{
\begin{array}{ll}
\! \! \! 1 \, &{\rm if} \, k=\ell , \\
\! \! \! \varrho_{k,\ell} \, & {\rm if} \, k \neq \ell ,
\end{array}
\right.
\end{equation}
with $ k,\ell=1,2,\ldots,L$, and where $\varrho_{k,\ell} \triangleq J_0^2 \big( \frac{2 \pi \left| k-\ell \right| W}{L-1} \big)$ is the correlation coefficient between $\left|h_k\right|^2$ and $\left|h_\ell\right|^2$ \cite[eq$.$~(7)]{wong2021fluid}, which can also be interpreted as the power correlation coefficient between $\left|h_k\right|$ and $\left|h_\ell\right|$.
\subsection{Pilot-Aided CE}
Considering pilot-aided CE, $N_{\text{p}}$ pilot symbols, each with energy $E_{\text{p}}=\varepsilon E_{\text{s}}$ (with $\varepsilon \in (0,1]$), are transmitted at the beginning of each coherence time frame~\cite{gifford2008antenna}. With ML estimation, the estimated $h_\ell$ is given by \cite[Sec$.$ II]{gifford2008antenna}
\beq
\hat{h}_\ell \triangleq \frac{\sum_{n=1}^{N_\text{p}}p_{\ell,n}}{\sqrt{E_{\text{p}}}N_{\text{p}}} = h_\ell+e_\ell,
\label{eq:estimations}
\eeq
where $p_{\ell,n}$ is the $n$-th pilot symbol received over the $\ell$-th FAS port and $e_\ell\sim\mathcal{C}\mathcal{N}\left(0,2\sigma_{\text{ce}}^2\right)$ is the corresponding CE error, with $\sigma_{\text{ce}}^2\triangleq \sigma_{\text{n}}^2/\left(2N_{\text{p}}E_{\text{p}}\right)$. Analyzing the performance of the considered $L$-port FAS under correlated Nakagami-$m$ fading with imperfect CSI requires analytical expressions for the joint statistics of $\big\{ |\hat{h}_\ell|^2 \big\}$ or functions thereof~\cite{simon2005digital}. In the following section, we present a tight approximation to these joint statistics when each $m_\ell$ is IoHI, together with exact expressions for the special case of Rayleigh fading, i.e., for $m_\ell=1$, $\forall$$\ell=1,2,\ldots,L$.
\section{Joint Statistics of the Estimated Channels} \label{Sec:Statistics}
Correlated Nakagami-$m$ RVs with an arbitrary power CM $\boldsymbol{\Sigma}$ can be generated from correlated Gaussian RVs with CM $\sqrt{\boldsymbol{\Sigma}}$ \cite{alexandropoulos2017swc}. Following this approach, we model each Nakagami-$m$-distributed channel gain amplitude $\left|h_\ell\right|$ with IoHI $m_\ell$ as
\beq
\left|h_\ell\right| \triangleq \left(\sum_{i=1}^{2m_\ell} x_{i,\ell}^2\right)^{1/2},
\label{eq:amplitude_model_h_ell}
\eeq
with $x_{i,\ell}\sim\mathcal{N}\left(0,\sigma_{\textrm{h}}^2/m_\ell\right)$, $\forall$$i=1,2,\ldots,m_\ell$. Clearly, one straightforwardly obtains $\mathbb{E}[|h_\ell|^2]=2 \sigma_{\textrm{h}}^2$, $\forall \ell$. We further adopt a similar approach to \cite{yacoub2005nakagami} to express each actual channel gain $h_\ell$ appearing in \eqref{eq:estimations} as a complex RV. Specifically, by considering that each Nakagami-$m$ RV $\left|h_\ell\right|$ is independent of the random phase $\phi_\ell$ with PDF given by \cite[eq$.$~(3)]{yacoub2005nakagami}, we generate the $\ell$-th actual channel gain as $h_\ell \triangleq y_{1,\ell} + \jmath y_{2,\ell}$, with
\beq
y_{q,\ell}\!\triangleq\!\left(\sum_{i=1}^{2m_\ell}\!x_{i,\ell}^2\right)^{1/2}\cos\bigg((q\!-\!1)\frac{\pi}{2}\!-\!(-1)^q\phi_\ell\bigg), \quad q=1,2.
\label{eq:model_h_ell}
\eeq  
The latter generation approach for $\{h_\ell\}$ easily leads to the Nakagami-$m$-distributed amplitude expression given by \eqref{eq:amplitude_model_h_ell} with $\mathbb{E}[|h_\ell|^2]=2 \sigma_{\textrm{h}}^2$, $\forall \ell$. Then, we substitute \eqref{eq:model_h_ell} into \eqref{eq:estimations} to express each estimated channel gain $\hat{h}_\ell$ as
\beq
\hat{h}_\ell \triangleq \left( y_{1,\ell} + {\rm Re} \left( e_\ell \right) \right)
+ \jmath \left( y_{2,\ell} + {\rm Im} \left( e_\ell \right) \right),
\label{eq:h_hat_Nakagami}
\eeq
with ${\rm Re}(e_\ell)$, ${\rm Im}(e_\ell)\!\!\sim\!\!\mathcal{N}\left(0,\sigma^2_{\text{ce}}\right)$. It is evident from \eqref{eq:h_hat_Nakagami} that neither the real nor the imaginary part of $\hat{h}_\ell$ is a Gaussian RV. This fact complicates the computation of the marginal distributions of $\big\{|\hat{h}_\ell|^2\big\}$ as well as their joint statistics, which are needed for the performance analysis of the $L$-port FAS receiver over the considered Nakagami-$m$ channel model.

\subsection{Marginal Distribution}

To overcome the above-mentioned difficulty and obtain mathematically tractable expressions for the distributions of the gains, we propose to approximate each $|\hat{h}_\ell|^2$ obtained from \eqref{eq:h_hat_Nakagami} by a Gamma RV, as shown in the following proposition.

\begin{proposition}
\label{Proposition:Gamma_Approximation}
Consider Nakagami-$m$-distributed RVs $\big\{|h_\ell|\big\}$ with IoHI $m_\ell$, $\forall \ell$, generated using \eqref{eq:amplitude_model_h_ell} and \eqref{eq:model_h_ell}, and complex-Gaussian-distributed CE errors $\{e_\ell\}$ as defined below \eqref{eq:estimations}. Each $|\hat{h}_\ell |^2$ can be tightly approximated by a Gamma RV having PDF
\beq
f_{|\hat{h}_\ell|^2} (x)
\triangleq \frac{x^{\kappa_\ell-1}}{\Gamma(\kappa_\ell)\theta_\ell^{\kappa_\ell}}
\exp \left(-\frac{x}{\theta_\ell} \right), \quad x \geq 0,
\label{eq:pdf_approx}
\eeq
with shape and scale parameters obtained using the method of moment matching as
\beq
\kappa_\ell \!\triangleq\! \frac{\mathbb{E}^2[|\hat{h}_\ell|^2]}{\mathbb{E}[|\hat{h}_\ell|^4]\!-\!\mathbb{E}^2[|\hat{h}_\ell|^2]},\quad
\theta_\ell \!\triangleq\! \frac{\mathbb{E}[|\hat{h}_\ell|^4]\!-\!\mathbb{E}^2[|\hat{h}_\ell|^2]}{\mathbb{E}[|\hat{h}_\ell|^2]} ,
\label{eq:Gamma_Approximation_Parameters}
\eeq
respectively. In the latter expression, the mean and second moment of $|\hat{h}_\ell|^2$ are obtained utilizing the statistical independence of the channel gains and the CE noises as
\begin{align}
\mathbb{E} [|\hat{h}_\ell|^2 ] & = 2 \left( \sigma_{\textrm{h}}^2 + \sigma^2_{\text{ce}} \right) , \\
\mathbb{E} [|\hat{h}_\ell|^4] & = 4 \sigma_{\textrm{h}}^4 \left( m_{\ell} +1\right)/m_\ell
+ 8 \sigma_{\text{ce}}^4 + 16 \sigma_{\textrm{h}}^2 \sigma_{\text{ce}}^2 ,
\label{eq:variance}
\end{align}
respectively.
\end{proposition}

\begin{IEEEproof}
Substituting \eqref{eq:model_h_ell} into \eqref{eq:estimations} results in the expression $|\hat{h}_\ell|^2=|h_\ell|^2+|e_\ell|^2+2\left(y_{1,\ell}{\rm Re}(e_\ell)+y_{2,\ell}{\rm Im}(e_\ell)\right)$. Furthermore, the statistical independence between each actual channel gain and its corresponding CE error directly yields the expression for $\mathbb{E}[|\hat{h}_\ell|^2]$. The second moment of $|\hat{h}_\ell|^2$ can similarly be expressed as
\begin{align}
\hspace{-2mm} \mathbb{E}[|\hat{h}_\ell|^4] = \ & \mathbb{E} [|h_\ell|^4] + \mathbb{E} [|e_\ell|^4]
+ 2\mathbb{E} [|h_\ell|^2] \mathbb{E}[|e_\ell|^2] \nn \\
& + 4 \mathbb{E} [y_{1,\ell}^2] \mathbb{E} [{\rm Re}^2(e_\ell)]
+ 4 \mathbb{E} [y_{2,\ell}^2] \mathbb{E} [{\rm Im}^2(e_\ell)] .
\label{eq:fourth_moment}
\end{align}
Note that $\mathbb{E} [|h_\ell|^4]$ is the second moment of the Gamma RV $|h_\ell|^2$ and is obtained as the first term of the result in \eqref{eq:variance}. Similarly, $\mathbb{E} [|e_\ell|^4]$ is the second moment of the exponentially distributed RV $|e_\ell|^2$ and is obtained as $8 \sigma_{\text{ce}}^4$. The computation of $\mathbb{E} [y_{1,\ell}^2]$ and $\mathbb{E} [y_{2,\ell}^2]$ requires the PDF of $\phi_\ell$, which is given by \cite[eq$.$~(3)]{yacoub2005nakagami}
\begin{equation}\label{eq:phi_pdf}
f_{\phi_\ell} (\theta)
\triangleq \frac{\Gamma(m_{\ell}) |\sin (2 \theta)|^{m_{\ell}-1}}{2^{m_{\ell}} \Gamma^2 (m_{\ell}/2)}.
\end{equation}
Thus, capitalizing on the statistical independence between $\phi_\ell$ and $x_{i,\ell}$, $\forall$$i=1,2,\ldots,m_\ell$, we have
\begin{align}
\mathbb{E} [y_{1,\ell}^2]
&= \mathbb{E} \left[ \left(\sum_{i=1}^{2m_\ell}\!x_{i,\ell}^2\right) \cos^2 (\phi_{\ell}) \right]
= 2 \sigma_{\textrm{h}}^2 \mathbb{E} \left[ \cos^2 (\phi_{\ell}) \right] \nn \\
& = \frac{2 \sigma_{\textrm{h}}^2 \Gamma (m_{\ell})}{2^{m_{\ell}} \Gamma^2 (m_{\ell}/2)}
\int_0^{2 \pi} \cos^2 (\theta) |\sin(2\theta)|^{m_{\ell}-1} {\rm d} \theta \nn \\
& \stackrel{(a)}{=} \frac{4 \sigma_{\textrm{h}}^2 \Gamma (m_{\ell})}{\Gamma^2 (m_{\ell}/2)}
\int_0^{\pi/2} \sin^{m_{\ell}-1} (\theta )\cos^{m_{\ell}+1} (\theta) {\rm d} \theta ,
\label{eq:exp_cal_y1}
\end{align}
where step $(a)$ is obtained by using the identity $\sin(2 \theta)=2 \sin (\theta) \cos (\theta)$ and by exploiting the symmetry of the absolute sinusoidal functions in the four quadrants. Furthermore, using the definitions $B(p/2,q/2) \triangleq 2 \int_0^{\pi/2} \sin^{p-1} \theta \cos^{q-1} \theta {\rm d} \theta$ and $B(x,y)\triangleq\Gamma(x) \Gamma(y)/\Gamma(x+y)$ in \eqref{eq:exp_cal_y1}, followed by algebraic simplifications, results in $\mathbb{E} [y_{1,\ell}^2] = \sigma_{\textrm{h}}^2$, $\forall \ell$. Similarly, we have
\begin{align}
\mathbb{E} [y_{2,\ell}^2]
&= \mathbb{E} \left[ \left(\sum_{i=1}^{2m_\ell}\!x_{i,\ell}^2\right) \sin^2 (\phi_{\ell}) \right]
= 2 \sigma_{\textrm{h}}^2 \mathbb{E} \left[ \sin^2 (\phi_{\ell}) \right] \nn \\
& = \frac{4 \sigma_{\textrm{h}}^2 \Gamma (m_{\ell})}{\Gamma^2 (m_{\ell}/2)}
\int_0^{\pi/2} \sin^{m_{\ell}+1} (\theta) \cos^{m_{\ell}-1} (\theta) {\rm d} \theta ,
\label{eq:exp_cal_y2}
\end{align}
which can be simplified using the same approach as in the computation of $\mathbb{E} [y_{1,\ell}^2]$, resulting in $\mathbb{E} [y_{2,\ell}^2] = \sigma_{\textrm{h}}^2$. The proof is completed by utilizing the expressions for the variances of the channel gains and the CE errors.
\end{IEEEproof}

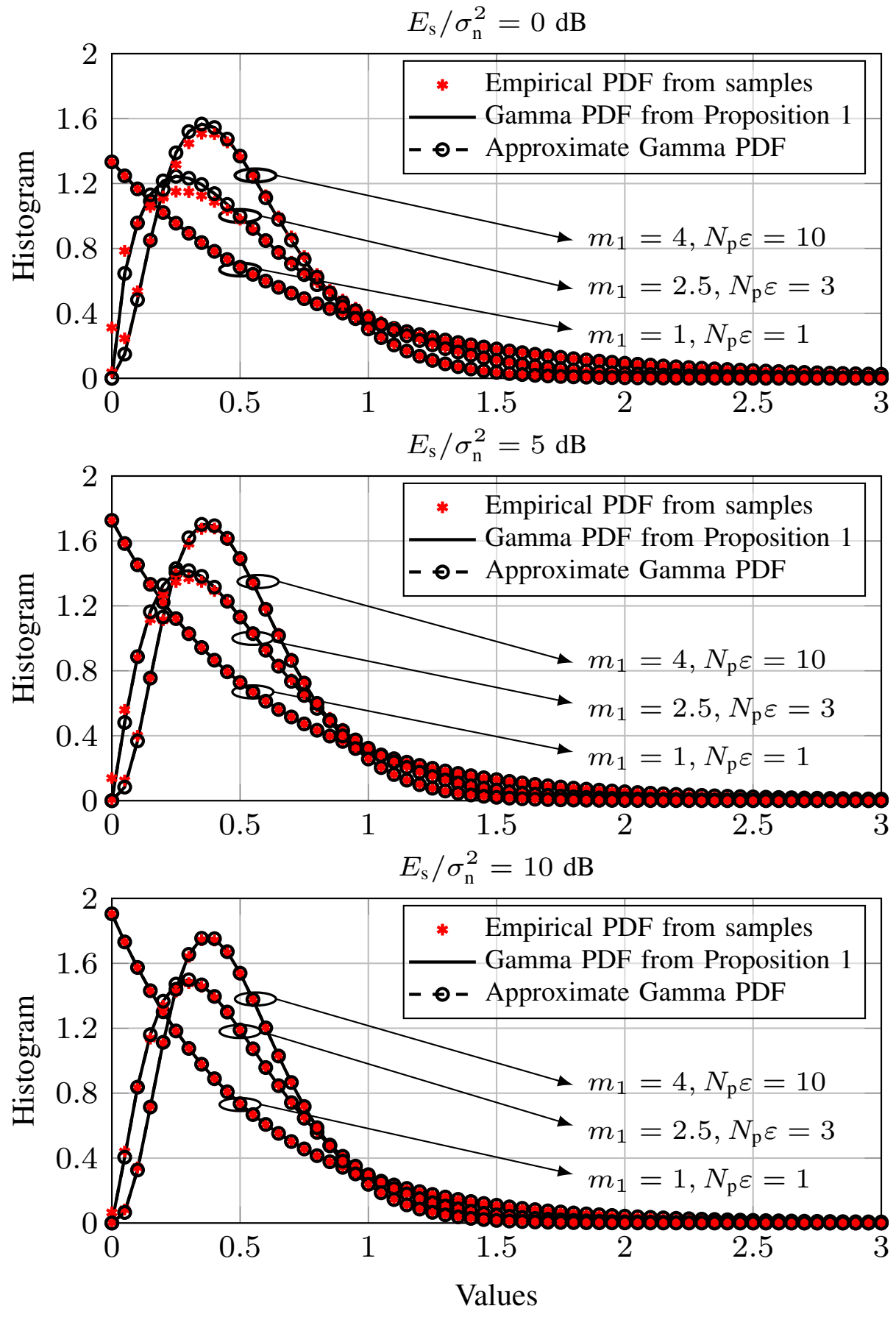
\begin{figure}[!t]
\centering
\pgfdeclarelayer{background}
\pgfdeclarelayer{foreground}
\pgfsetlayers{background,main,foreground}

\begin{tikzpicture}

\begin{groupplot}[
    group style={
        group size=1 by 3,
        vertical sep=25pt,
        y descriptions at=edge left,
    },
    width=8.5cm,
    height=4.5cm,
    xlabel near ticks,
    ylabel near ticks,
    ylabel style={yshift=-3pt},
    label style={font=\footnotesize},
    ticklabel style={font=\footnotesize},
    legend style={
        at={(0.98,0.98)},
        anchor=north east,
        font=\scriptsize,
        inner sep=1pt,
        row sep=-2pt,
        fill opacity=0.5,
        draw opacity=1,
        text opacity=1
    },
    legend cell align=left,
    grid=major,
]
\nextgroupplot[
    xlabel={},
    ylabel={Histogram},
    xmin=0, xmax=3,
    xtick={1,0.5,...,3},
    ymin=0, ymax=2,
    ytick={0,0.4,...,2},
    title={$E_{\text{s}}/\sigma_{\text{n}}^2=0$ dB},
    title style={font=\scriptsize,yshift=-2mm},
]

\addplot[red, thick, only marks, mark=asterisk, mark size=1.5pt]
table [x=range, y=pdf_hhat, col sep=comma] {Figures/files_txt/plot1/1m_1Npe_0SNRdB.txt};
\addlegendentry{Empirical PDF from samples};

\addplot[black, thick]
table [x=range, y=analyticalPDF_hhat_Prop1, col sep=comma] {Figures/files_txt/plot1/1m_1Npe_0SNRdB.txt};
\addlegendentry{Gamma PDF from Proposition 1};

\addplot[black, thick, dashed, mark=o, mark options={solid}, mark size=1.5pt]
table [x=range, y=analyticalPDF_hhat_Approx, col sep=comma] {Figures/files_txt/plot1/1m_1Npe_0SNRdB.txt};
\addlegendentry{Approximate Gamma PDF};

\addplot[red, thick, only marks, mark=asterisk, mark size=1.5pt]
table [x=range, y=pdf_hhat, col sep=comma] {Figures/files_txt/plot1/2.5m_3Npe_0SNRdB.txt};

\addplot[black, thick]
table [x=range, y=analyticalPDF_hhat_Prop1, col sep=comma] {Figures/files_txt/plot1/2.5m_3Npe_0SNRdB.txt};

\addplot[black, thick, dashed, mark=o, mark options={solid}, mark size=1.5pt]
table [x=range, y=analyticalPDF_hhat_Approx, col sep=comma] {Figures/files_txt/plot1/2.5m_3Npe_0SNRdB.txt};

\addplot[red, thick, only marks, mark=asterisk, mark size=1.5pt]
table [x=range, y=pdf_hhat, col sep=comma] {Figures/files_txt/plot1/4m_10Npe_0SNRdB.txt};

\addplot[black, thick]
table [x=range, y=analyticalPDF_hhat_Prop1, col sep=comma] {Figures/files_txt/plot1/4m_10Npe_0SNRdB.txt};

\addplot[black, thick, dashed, mark=o, mark options={solid}, mark size=1.5pt]
table [x=range, y=analyticalPDF_hhat_Approx, col sep=comma] {Figures/files_txt/plot1/4m_10Npe_0SNRdB.txt};

\draw[thick] (0.5,0.67) ellipse (0.08 and 0.04);
\draw[thick] (0.5,1) ellipse (0.08 and 0.04);
\draw[thick] (0.56,1.25) ellipse (0.08 and 0.04);

\begin{scope}[>=latex]
\draw[->] (0.58,0.67) -- (1.8,0.3);
\draw[->] (0.58,1) -- (1.8,0.55);
\draw[->] (0.64,1.25) -- (1.8,0.85);
\end{scope}

\node[black, font=\scriptsize, anchor=west] at (1.8,0.25) {$m_1 = 1, N_{\text{p}} \varepsilon=1$};
\node[black, font=\scriptsize, anchor=west] at (1.8,0.55) {$m_1 = 2.5, N_{\text{p}} \varepsilon=3$};
\node[black, font=\scriptsize, anchor=west] at (1.8,0.85) {$m_1 = 4, N_{\text{p}} \varepsilon=10$};
\nextgroupplot[
    xlabel={},
    ylabel={Histogram},
    xmin=0, xmax=3,
    xtick={1,0.5,...,3},
    ymin=0, ymax=2,
    ytick={0,0.4,...,2},
    title={$E_{\text{s}}/\sigma_{\text{n}}^2=5$ dB},
    title style={font=\scriptsize,yshift=-2mm},
]
\addplot[red, thick, only marks, mark=asterisk, mark size=1.5pt]
table [x=range, y=pdf_hhat, col sep=comma] {Figures/files_txt/plot1/1m_1Npe_5SNRdB.txt};
\addlegendentry{Empirical PDF from samples};

\addplot[black, thick]
table [x=range, y=analyticalPDF_hhat_Prop1, col sep=comma] {Figures/files_txt/plot1/1m_1Npe_5SNRdB.txt};
\addlegendentry{Gamma PDF from Proposition 1};

\addplot[black, thick, dashed, mark=o, mark options={solid}, mark size=1.5pt]
table [x=range, y=analyticalPDF_hhat_Approx, col sep=comma] {Figures/files_txt/plot1/1m_1Npe_5SNRdB.txt};
\addlegendentry{Approximate Gamma PDF};

\addplot[red, thick, only marks, mark=asterisk, mark size=1.5pt]
table [x=range, y=pdf_hhat, col sep=comma] {Figures/files_txt/plot1/2.5m_3Npe_5SNRdB.txt};

\addplot[black, thick]
table [x=range, y=analyticalPDF_hhat_Prop1, col sep=comma] {Figures/files_txt/plot1/2.5m_3Npe_5SNRdB.txt};

\addplot[black, thick, dashed, mark=o, mark options={solid}, mark size=1.5pt]
table [x=range, y=analyticalPDF_hhat_Approx, col sep=comma] {Figures/files_txt/plot1/2.5m_3Npe_5SNRdB.txt};

\addplot[red, thick, only marks, mark=asterisk, mark size=1.5pt]
table [x=range, y=pdf_hhat, col sep=comma] {Figures/files_txt/plot1/4m_10Npe_5SNRdB.txt};

\addplot[black, thick]
table [x=range, y=analyticalPDF_hhat_Prop1, col sep=comma] {Figures/files_txt/plot1/4m_10Npe_5SNRdB.txt};

\addplot[black, thick, dashed, mark=o, mark options={solid}, mark size=1.5pt]
table [x=range, y=analyticalPDF_hhat_Approx, col sep=comma] {Figures/files_txt/plot1/4m_10Npe_5SNRdB.txt};

\draw[semithick] (0.55,0.67) ellipse (0.08 and 0.04);
\draw[semithick] (0.55,1) ellipse (0.08 and 0.04);
\draw[semithick] (0.57,1.35) ellipse (0.08 and 0.04);

\begin{scope}[>=latex]
\draw[->] (0.63,0.67) -- (1.8,0.3);
\draw[->] (0.63,0.98) -- (1.8,0.6);
\draw[->] (0.65,1.35) -- (1.8,0.85);
\end{scope}

\node[black, font=\scriptsize, anchor=west] at (1.8,0.25) {$m_1 = 1, N_{\text{p}} \varepsilon=1$};
\node[black, font=\scriptsize, anchor=west] at (1.8,0.55) {$m_1 = 2.5, N_{\text{p}} \varepsilon=3$};
\node[black, font=\scriptsize, anchor=west] at (1.8,0.85) {$m_1 = 4, N_{\text{p}} \varepsilon=10$};
\nextgroupplot[
    xlabel={Values},
    ylabel={Histogram},
    xmin=0, xmax=3,
    xtick={1,0.5,...,3},
    ymin=0, ymax=2,
    ytick={0,0.4,...,2},
    title={$E_{\text{s}}/\sigma_{\text{n}}^2=10$ dB},
    title style={font=\scriptsize,yshift=-2mm},
]
\addplot[red, thick, only marks, mark=asterisk, mark size=1.5pt]
table [x=range, y=pdf_hhat, col sep=comma] {Figures/files_txt/plot1/1m_1Npe_10SNRdB.txt};
\addlegendentry{Empirical PDF from samples};

\addplot[black, thick]
table [x=range, y=analyticalPDF_hhat_Prop1, col sep=comma] {Figures/files_txt/plot1/1m_1Npe_10SNRdB.txt};
\addlegendentry{Gamma PDF from Proposition 1};

\addplot[black, thick, dashed, mark=o, mark options={solid}, mark size=1.5pt]
table [x=range, y=analyticalPDF_hhat_Approx, col sep=comma] {Figures/files_txt/plot1/1m_1Npe_10SNRdB.txt};
\addlegendentry{Approximate Gamma PDF};

\addplot[red, thick, only marks, mark=asterisk, mark size=1.5pt]
table [x=range, y=pdf_hhat, col sep=comma] {Figures/files_txt/plot1/2.5m_3Npe_10SNRdB.txt};

\addplot[black, thick]
table [x=range, y=analyticalPDF_hhat_Prop1, col sep=comma] {Figures/files_txt/plot1/2.5m_3Npe_10SNRdB.txt};

\addplot[black, thick, dashed, mark=o, mark options={solid}, mark size=1.5pt]
table [x=range, y=analyticalPDF_hhat_Approx, col sep=comma] {Figures/files_txt/plot1/2.5m_3Npe_10SNRdB.txt};

\addplot[red, thick, only marks, mark=asterisk, mark size=1.5pt]
table [x=range, y=pdf_hhat, col sep=comma] {Figures/files_txt/plot1/4m_10Npe_10SNRdB.txt};

\addplot[black, thick]
table [x=range, y=analyticalPDF_hhat_Prop1, col sep=comma] {Figures/files_txt/plot1/4m_10Npe_10SNRdB.txt};

\addplot[black, thick, dashed, mark=o, mark options={solid}, mark size=1.5pt]
table [x=range, y=analyticalPDF_hhat_Approx, col sep=comma] {Figures/files_txt/plot1/4m_10Npe_10SNRdB.txt};

\draw[semithick] (0.5,0.73) ellipse (0.08 and 0.04);
\draw[semithick] (0.5,1.18) ellipse (0.08 and 0.04);
\draw[semithick] (0.56,1.38) ellipse (0.08 and 0.04);

\begin{scope}[>=latex]
\draw[->] (0.58,0.73) -- (1.8,0.3);
\draw[->] (0.58,1.18) -- (1.8,0.6);
\draw[->] (0.64,1.38) -- (1.8,0.85);
\end{scope}

\node[black, font=\scriptsize, anchor=west] at (1.8,0.25) {$m_1 = 1, N_{\text{p}} \varepsilon=1$};
\node[black, font=\scriptsize, anchor=west] at (1.8,0.55) {$m_1 = 2.5, N_{\text{p}} \varepsilon=3$};
\node[black, font=\scriptsize, anchor=west] at (1.8,0.85) {$m_1 = 4, N_{\text{p}} \varepsilon=10$};
\end{groupplot}
\end{tikzpicture}\vspace{-2mm}
\caption{Histogram of the RV $|\hat{h}_1|^2$ for $E_{\text{s}}/\sigma_{\text{n}}^2 \in \{0,5,10\}$~dB and different values of $m_1$ and $N_{\text{p}}\varepsilon$.}
\vspace{-3mm}
\label{f1}
\end{figure}
Fig.~\ref{f1} illustrates the histogram of $|\hat{h}_1|^2$ for $E_{\text{s}}/\sigma_{\text{n}}^2 \in \{0,5,10\}$~dB and different values of $m_1$ and the energy devoted to pilot symbols $N_{\text{p}}\varepsilon$. The same figure also presents the curves for the Gamma PDF of Proposition~\ref{Proposition:Gamma_Approximation} together with another Gamma PDF termed as ``Approximate Gamma PDF'', which has $\theta_1$ as its scale parameter and the shape parameter $\hat{m}_1$ given by the smallest closest IoHI to $\kappa_1$. Both approximations for the PDF of $|\hat{h}_1|^2$ are tight for all considered values for $m_1$ and $N_{\text{p}} \varepsilon$, and become even tighter with increasing $E_{\text{s}}/\sigma_{\text{n}}^2$ values. Following the tightness of the ``Approximate Gamma PDF'', as depicted in Fig.~\ref{f1}, we propose to alternatively model each $|\hat{h}_\ell|$ as a Nakagami-$m$-distributed RV with IoHI fading parameter $\hat{m}_\ell$ and average fading power $\hat{m}_\ell\theta_\ell$, and consequently generate it from Gaussian RVs as in \eqref{eq:amplitude_model_h_ell} by just replacing $m_\ell$ with $\hat{m}_\ell$ and $x_{i,\ell}$, $\forall$$i=1,2,\ldots,\hat{m}_\ell$, with $\tilde{x}_{i,\ell}\sim\mathcal{N}\left(0,\theta_\ell/2\right)$.

\subsection{Multivariate Distribution}
\subsubsection{Correlation Matrix}
We next present an analytical expression for the correlation coefficient between $|\hat{h}_k|$ and $|\hat{h}_\ell|$, $\forall$$k\neq\ell$, which is essential for obtaining analytical expressions for the joint statistics of $\big\{|\hat{h}_\ell|\big\}$ and functions thereof.
\begin{proposition}
\label{Proposition:Approximated_Correlation_Coefficient_Nakagami}
Consider two arbitrarily correlated Nakagami-$m$ RVs $|h_k|$ and $|h_\ell|$ obtained using \eqref{eq:amplitude_model_h_ell} from Gaussian RVs $x_{i,k}$, $\forall$$i=1,2,\ldots,m_k$, and $x_{i,\ell}$, $\forall$$i=1,2,\ldots,m_\ell$, respectively, for IoHI $m_k$ and $m_\ell$. Also, consider two complex-Gaussian-distributed CE errors $e_k$ and $e_\ell$ as defined below \eqref{eq:estimations}. By approximating $|\hat{h}_k|$ obtained from \eqref{eq:h_hat_Nakagami} with a Nakagami-$m$ RV generated from Gaussian RVs $\tilde{x}_{i,k}$, $\forall i=1,2,\ldots,\hat{m}_k$, with parameters $\hat{m}_k$ and $\hat{m}_k\theta_k$, and similarly $|\hat{h}_\ell|$ with another Nakagami-$m$ RV generated from $\tilde{x}_{i,\ell}$, $\forall i=1,2,\ldots,\hat{m}_\ell$, with parameters $\hat{m}_\ell$ and $\hat{m}_\ell\theta_\ell$, the Gaussian correlation coefficient between $\tilde{x}_{i,k}$ and $\tilde{x}_{i,\ell}$ for $k\neq\ell$, $\forall i=1,2,\ldots,2\min(\hat{m}_k,\hat{m}_\ell)$, is given by
\beq
\hat{\rho}_{k,\ell} \triangleq 2\sigma_{h}^2\left(m_km_\ell\theta_k\theta_\ell\right)^{-1/2}\sqrt{\varrho_{k,\ell}},
\label{eq:gaussian_correlation_coefficient_nakagami}
\eeq
where $\sqrt{\varrho_{k,\ell}}$ is the Gaussian correlation coefficient between $x_{i,k}$ and $x_{i,\ell}$ for $k\neq\ell$, $\forall i=1,2,\ldots,2\min(m_k,m_\ell)$.
\end{proposition}

\begin{IEEEproof}
Summing the independent Gaussian RVs $x_{i,k}$ and $z_k\sim\mathcal{N}\left(0,\theta_k/2-\sigma_{h}^2/m_k\right)$ yields $\tilde{x}_{i,k}$. Similarly, one derives $\tilde{x}_{i,\ell}$ from $x_{i,\ell}$ and $z_\ell\sim\mathcal{N}\left(0,\theta_\ell/2-\sigma_{h}^2/m_\ell\right)$. Due to the statistical independence between $x_{i,k}$ and $z_k$, $x_{i,\ell}$ and $z_\ell$, and making the reasonable assumption that $z_k$ and $z_\ell$ are independent, the covariance between $\hat{x}_{i,k}$ and $\hat{x}_{i,\ell}$ coincides with the covariance of $x_{i,k}$ and $x_{i,\ell}$. The proof is completed by combining the latter derivation together with the expressions for the variances of $x_{i,k}$, $x_{i,\ell}$, $\tilde{x}_{i,k}$, and $\tilde{x}_{i,\ell}$ into the definition of the Gaussian correlation coefficient \cite[eq$.$~(9C$.$2)]{simon2005digital}.
\end{IEEEproof}

Proposition~\ref{Proposition:Approximated_Correlation_Coefficient_Nakagami} reveals that estimating correlated Nakagami-$m$ channel gains using pilot symbols and the ML estimator described in \eqref{eq:estimations} leads approximately to correlated Nakagami-$m$ channel gain estimates. In fact, $\big\{|\hat{h}_\ell|\big\}$ are arbitrarily correlated since they are generated from arbitrarily correlated Gaussian RVs, as shown in \eqref{eq:gaussian_correlation_coefficient_nakagami}. Interestingly, the existence of CE errors results in estimated channels that are less correlated than the actual ones, as will be shown next. Moreover, from \eqref{eq:Gamma_Approximation_Parameters} and \eqref{eq:variance}, we have $\kappa_{\ell}=(\sigma_{\textrm{h}}^2+\sigma_{\text{ce}}^2)^2/(\sigma_{\textrm{h}}^4/m_{\ell} + \sigma_{\text{ce}}^4 + 2 \sigma_{\textrm{h}}^2\sigma_{\text{ce}}^2)$ and $\kappa_k\theta_k=\kappa_\ell\theta_\ell=\mathbb{E} [ |\hat{h}_{\ell}|^2 ]=2(\sigma_{\textrm{h}}^2 + \sigma^2_{\text{ce}})$. Using the fact that $m_{\ell} \geq 1$, $\forall \ell$, we have $ \sigma_{\textrm{h}}^4 + m_{\ell} (\sigma_{\text{ce}}^4 + 2 \sigma_{\textrm{h}}^2 \sigma_{\text{ce}}^2) \geq (\sigma_{\textrm{h}}^2+\sigma_{\text{ce}}^2)^2$, resulting in $m_\ell\geq\kappa_\ell$ and $m_k\geq\kappa_k$. This implies $m_k \theta_k \geq \kappa_k \theta_k$ and $m_\ell \theta_\ell \geq \kappa_\ell \theta_\ell$, which further yields
\beq
0<2\sigma_{h}^2\left(m_km_\ell\theta_k\theta_\ell\right)^{-1/2}\leq \sigma_{\textrm{h}}^2/(\sigma_{\textrm{h}}^2 + \sigma_{\text{ce}}^2) \leq 1.
\label{eq:parameter}
\eeq
The above, after substitution into \eqref{eq:gaussian_correlation_coefficient_nakagami}, gives $\hat{\rho}_{k,\ell}\leq\sqrt{\varrho_{k,\ell}}$. Clearly, as $\sigma_{\text{ce}}^2$ increases (i$.$e$.$, $E_{\text{s}}/\sigma_{\text{n}}^2$ and/or $N_{\text{p}}\varepsilon$ decrease), the gap between $\hat{\rho}_{k,\ell}$ and $\sqrt{\varrho_{k,\ell}}$ widens. Conversely, for $\sigma_{\text{ce}}^2\rightarrow0$ (i$.$e$.$, almost perfect CE), we have $\hat{\rho}_{k,\ell}\rightarrow\sqrt{\varrho_{k,\ell}}$.

\subsubsection{Joint Statistics}
Capitalizing on the approximation approach of Proposition~\ref{Proposition:Approximated_Correlation_Coefficient_Nakagami}, analytical expressions for the joint statistics of the estimated channel gains can be directly obtained from the expressions for the joint statistics of the actual Nakagami-$m$ channel gains available in the literature (see, e$.$g$.$, \cite{win2000mrc, 4786505, alexandropoulos2017swc}). Specifically, one just needs to perform the following replacements in the available expressions: \textit{i})~$\mathbb{E}[|h_\ell|^2]$ with $\mathbb{E}[|\hat{h}_\ell|^2]$, $\forall \ell$; and \textit{ii})~the Gaussian CM $\sqrt{\boldsymbol{\Sigma}}$ of $\big\{|h_\ell|\big\}$ with the symmetric positive definite Gaussian CM $\hat{\mathbf{\Sigma}}_{\mathbf{G}}\in\mathbb{R}^{L\times L}_+$ defined as
\beq
\! \! \! \! [\hat{\mathbf{\Sigma}}_{\mathbf{G}}]_{k,\ell} \triangleq
\left\{
\begin{array}{ll}
\! \! \! 1 \, &{\rm if} \, k=\ell , \\
\! \! \! \hat{\rho}_{k,\ell} \, & {\rm if} \, k \neq \ell ,
\end{array}
\right.
\label{eq:CM_Est}
\eeq
with $ k,\ell=1,2,\ldots,L$.

It is noted that the analysis carried out above is valid for any arbitrary power CM, and thus can be directly applied to the performance analysis of the considered GSC-based FAS receiver using the power CM given in~\eqref{eq:CM_FAS}. In particular, to aid the SEP characterization, we consider the ordered estimated channel gain vector $\mathbf{\hat{h}}_{[L]} \triangleq [\hat{h}_{[1]},\hat{h}_{[2]},\ldots,\hat{h}_{[L]}]$, with $|\hat{h}_{[1]}| > |\hat{h}_{[2]}| > \ldots > |\hat{h}_{[L]}|$. Using the virtual branch technique presented in \cite[Sec$.$ II-B]{win2000mrc}, the characteristic function (CF) of $\|\mathbf{\hat{h}}_{[\tilde{L}]} \|^2$ for $\tilde{L}<L$ can be expressed using the joint statistics of the estimated channels presented above and \cite[eq. (15)]{gifford2008antenna} as
\begin{align}
\! \! \!
&\Psi_{\|\mathbf{\hat{h}}_{[\tilde{L}]} \|^2} \left( \jmath \omega \right) \triangleq
\prod_{q=1}^{\tilde{K}}
\left(1-2[\boldsymbol{\lambda}]_q(\sigma_{\textrm{h}}^2+\sigma_{\text{ce}}^2) \jmath \omega \right)^{-\frac{[\boldsymbol{\mu}]_q}{2}} \nn \\
& \times \! \! \! \! \prod_{q=\tilde{K}+1}^{K} \! \! \! 
\left(1-\frac{2M[\boldsymbol{\lambda}]_q}{q} (\sigma_{\textrm{h}}^2+\sigma_{\text{ce}}^2) \jmath \omega \right)^{-\frac{[\boldsymbol{\mu}]_q}{2}} \! \! ,
\label{eq:cf_fas}
\end{align}
where $\boldsymbol{\lambda}\in\mathbb{R}_+^{1\times K}$ contains the $K$ distinct eigenvalues (sorted in descending order) of $\mathbf{C}\in\mathbb{R}^{{\rm N}\times {\rm N}}$, with ${\rm N}\triangleq\sum_{\ell=1}^{L}\,2\hat{m}_\ell$, with corresponding algebraic multiplicities included in $\boldsymbol{\mu}\in\mathbb{Z}_+^{1\times K}$ such that $\sum_{q=1}^{K}[\boldsymbol{\mu}]_q = {\rm N}$. The matrix $\mathbf{C}$ is obtained as $\mathbf{C}\triangleq\mathbf{c}\mathbf{c}^{\rm T}$, with $\mathbf{c}\triangleq[\mathbf{b}_1^{\rm T},\mathbf{b}_2^{\rm T},\ldots,\mathbf{b}_{L}^{\rm T}]^{\rm T} \in\mathbb{R}^{{\rm N}\times 1}$ and $\mathbf{b}_\ell\triangleq[b_{\ell,1},b_{\ell,2},\ldots,b_{\ell,2\,\hat{m}_\ell}]^{\rm T}\in\mathbb{R}^{2\hat{m}_\ell\times 1}$. For any $\mathbf{b}_k$ and $\mathbf{b}_\ell$, we have
\beq
b_{k,j}\,b_{\ell,n} =
\left\{
\begin{array}{ll}
\! \! \! 1/\hat{m}_k \, &{\rm if}\,k=\ell\,{\rm and}\,j=n \\
\! \! \! \hat{\rho}_{k,\ell} \left(\hat{m}_k \hat{m}_\ell \right)^{-1/2} \, &{\rm if}\,k\neq\ell\,{\rm and}\,j=n \leq \xi_{k,\ell} \\
0 \, &{\rm otherwise}
\end{array}
\right.\!\!\!,
\label{eq:elements_of_C}
\eeq
with $\xi_{k,\ell}\triangleq2\min\left\{\hat{m}_k,\hat{m}_\ell\right\}$.

It is noted that, for $\hat{m}_\ell=\hat{m}$, $\forall \ell$, and $\tilde{L}=L$ (i.e., for the case of MRC-based reception), the CF of $\|\mathbf{\hat{h}}_L\|^2$ can be alternatively obtained by the simplified expression \cite[eq$.$~(11)]{lombardo1991mrc}.

\subsubsection{The Special Case of Rayleigh Distribution}
For the special case of correlated Rayleigh-distributed $\big\{|h_\ell|\big\}$, i.e., $m_\ell=1$, $\forall\ell$, an exact expression for the correlation coefficient $\hat{\rho}_{k,\ell}$ between the Gaussian RVs used to generate any two estimated channel gains $\hat{h}_k$ and $\hat{h}_\ell$, for $k\neq\ell$, can be derived as follows. Starting with $x_{i,\ell}\sim\mathcal{N}\left(0,\sigma_{\textrm{h}}^2\right)$, $i=1,2$, the $\ell$-th actual channel gain can be generated as $h_\ell = x_{1,\ell} + \jmath x_{2,\ell}$, indicating that each $h_\ell$ is a circularly symmetric complex Gaussian RV having the Rayleigh-distributed amplitude expression given by \eqref{eq:amplitude_model_h_ell} for $m_\ell=1$. By inserting the latter expression for $h_\ell$ into \eqref{eq:estimations} leads to $\hat{h}_\ell = \tilde{x}_{1,\ell} + \jmath \tilde{x}_{2,\ell}$, with $\tilde{x}_{1,\ell}\triangleq x_{1,\ell} + {\rm Re}(e_\ell)$ and $\tilde{x}_{2,\ell}\triangleq x_{2,\ell} + {\rm Im}(e_\ell)$. Since the actual channel gains are independent of the CE errors, $\tilde{x}_{1,\ell},\tilde{x}_{2,\ell}\sim\mathcal{N}\left(0,\sigma_{\textrm{h}}^2+\sigma^2_{\text{ce}}\right)$ is deduced, which reveals that each $|\hat{h}_\ell|$ is a Rayleigh-distributed RV with $\mathbb{E}[|\hat{h}_\ell|^2]=2(\sigma_{\textrm{h}}^2+\sigma^2_{\text{ce}})$.

\begin{proposition}
\label{Proposition:Approximated_Correlation_Coefficient_Rayleigh}
For Rayleigh-distributed $\big\{|h_\ell|\big\}$, i$.$e$.$, $m_\ell=1$, $\forall \ell$, the Gaussian correlation coefficient between $\tilde{x}_{i,k}$ and $\tilde{x}_{i,\ell}$, for $k\neq\ell$ and $i=1,2$, is given in closed form by
\begin{equation}
\hat{\rho}_{k,\ell} = \sigma_{\textrm{h}}^2/(\sigma_{\textrm{h}}^2 + \sigma_{\text{ce}}^2)\sqrt{\varrho_{k,\ell}} .
\label{eq:Gaussian_correlation_coefficient_Rayleigh}
\end{equation}
\end{proposition}

\begin{IEEEproof}
The proof follows similar steps as the proof of Proposition~\ref{Proposition:Approximated_Correlation_Coefficient_Nakagami}.  
\end{IEEEproof}
Substituting the right-hand side inequality of \eqref{eq:parameter} into \eqref{eq:Gaussian_correlation_coefficient_Rayleigh} results in $\hat{\rho}_{k,\ell}\leq\sqrt{\varrho_{k,\ell}}$. Analytical expressions for the joint statistics of $\big\{|\hat{h}_\ell|\big\}$ or functions thereof can be obtained as in the Nakagami-$m$ case (e.g., by setting $m_\ell=1$, $\forall\ell$, in \eqref{eq:cf_fas}).
\section{SEP Analysis}
\label{Sec:MRD}
In this section, using the previously derived joint statistics of the estimated Nakagami-$m$ channel gains, we present novel exact and asymptotic expressions for the SEP performance of a GSC-based FAS receiver with diverse modulation schemes.
Recall that, with GSC diversity, the receiver selects the best $\tilde{L}$ of the $L$ available ports (i.e., the top $\tilde{L}$ ports with the highest estimated channel gain amplitudes) and combines their received signals using MRC, prior to data detection using the ML rule. In particular, for the non-ideal estimates of the channel gains, the receiver collects the signals from the $\tilde{L}$ ports, cophases them, and applies appropriate weights to maximize the instantaneous SNR at the receiver's output. Thus, the instantaneous estimated SNR at the output of a GSC-based FAS receiver is given as $\gamma_{\rm FAS} \triangleq \|\mathbf{\hat{h}}_{[\tilde{L}]} \|^2 E_{\text{s}}/\sigma_{\text{n}}^2$.  We henceforth denote the considered diversity reception scheme as GSC$(\tilde{L},L)$.
\subsection{Exact Expressions}
\subsubsection{Non-Identical Fading Parameters}
Following the analysis in \cite{gifford2008antenna} and using the closed-form CF expression in~\eqref{eq:cf_fas}, an analytical expression for the SEP with $M$-PSK of the considered GSC$(\tilde{L},L)$-based FAS receiver is given by
\begin{align}
&P_{e,{\rm FAS}}^{\rm PSK} \triangleq \frac{1}{\pi} \int_0^{\frac{\pi (M-1)}{M}} \! \!
\prod_{q=1}^{\tilde{K}} \left( \frac{\sin^2 (\zeta)}
{\sin^2 (\zeta) + \eta_{\rm PSK} \left( \overline{\gamma} \right) [\boldsymbol{\lambda}]_q} \right)^{\frac{[\boldsymbol{\mu}]_q}{2}} \nn \\
& \times \! \! \! \prod_{q=\tilde{K}+1}^{K} \! \! \left( \frac{q \sin^2 \zeta}
{q \sin^2 (\zeta) + \tilde{K} \eta_{\rm PSK} \left( \overline{\gamma} \right) [\boldsymbol{\lambda}]_q} \right)^{\frac{[\boldsymbol{\mu}]_q}{2}} \! \! {\rm d} \zeta ,
\label{eq:sep_mpsk_fas}
\end{align}
with $\eta_{\rm PSK} \dn c_{\rm PSK} \overline{\gamma} N_{\text{p}} \varepsilon/(1/\overline{\gamma}+N_{\text{p}}\varepsilon+1)$ and $c_{\rm PSK} \dn \sin^2(\pi/M)$.
Similarly, an analytical expression for the SEP with $M$-QAM modulated symbols can be obtained as
\begin{align}
& P_{e, {\rm FAS}}^{\rm QAM}\triangleq \frac{1}{2\pi M}  \nn \\
& \times\! \! \left( \! \sum_{i=1}^{d_M} \omega_i^{\rm a} \! \! \int\limits_0^{\pi/2} \!
\prod_{q=1}^{\tilde{K}} \! \! \left( \! \frac{\sin^2 \left( \zeta \! + \! \frac{\pi}{4} \right)}
{\sin^2 \left(\zeta \! + \! \frac{\pi}{4} \right) \! + \! \eta_{i, {\rm QAM}} \left(\overline{\gamma}\right)[\boldsymbol{\lambda}]_q} \! \right)^{\! \! \frac{\! \![\boldsymbol{\mu}]_q}{2}} \right. \nn \\
& \times \! \! \! \prod_{q=\tilde{K}+1}^{K} \! \! \left( \frac{q \sin^2 \left( \zeta \! + \! \frac{\pi}{4} \right)}
{q \sin^2 \left( \zeta \! + \! \frac{\pi}{4} \right) \! + \! \tilde{K} \eta_{i,\rm QAM} \left( \overline{\gamma} \right) [\boldsymbol{\lambda}]_q} \right)^{\frac{\! \![\boldsymbol{\mu}]_q}{2}} \! \! {\rm d} \zeta \nn \\
& + \omega_i^{\rm b} \! \! \! \int\limits_0^{3\pi/4} \! \!
\prod_{q=1}^K \left(\frac{\sin^2 (\zeta)}
{\sin^2 (\zeta) + \eta_{i,{\rm QAM}} \left(\overline{\gamma}\right)[\boldsymbol{\lambda}]_q}\right)^{\! \! \frac{\! \! [\boldsymbol{\mu}]_q}{2}} \nn \\
& \left. \times \! \! \!
\prod_{q=\tilde{K}+1}^{K} \! \! \left( \! \frac{q \sin^2 (\zeta)}
{q \sin^2 (\zeta) \! + \! \tilde{K} \eta_{i,\rm QAM} \left( \overline{\gamma} \right) [\boldsymbol{\lambda}]_q} \! \right)^{\! \! \frac{[\boldsymbol{\mu}]_q}{2}} \! \! \! \! {\rm d} \zeta \right) \! ,
\label{eq:sep_mqam_fas}
\end{align}
where $\omega_{i}^{\rm a}$ and $\omega_{i}^{\rm b}$ depend on $M$, and their values for $M=\left\{4,16,64,256\right\}$ can be found in \cite[Table~I]{gifford2008antenna}. The finite summation in the latter expression is performed over the nonzero values of $\omega_{i}^{\rm a} $ and $\omega_{i}^{\rm b}$, and the included functions $\eta_{i,{\rm QAM}} \left(\overline{\gamma}\right)$, $\forall$$i=1,2,\ldots,d_M$, of the common actual average received SNR per branch $\overline{\gamma}$ are given by $\eta_{i,{\rm QAM}} \left(\overline{\gamma}\right) \dn c_{\rm QAM} \overline{\gamma} N_{\text{p}} \varepsilon/(1/\overline{\gamma}+N_{\text{p}}\varepsilon+\psi_i)$, with $c_{\rm QAM} \triangleq 3/(2(M-1))$ and where $\psi_i$ depends on $M$ \cite[Table~I]{gifford2008antenna}. Moreover, $\big\{ [\boldsymbol{\lambda}]_q \big\}$ are the eigenvalues of \eqref{eq:CM_Est} obtained using \eqref{eq:gaussian_correlation_coefficient_nakagami} and \eqref{eq:CM_FAS}. 

\subsubsection{Identical Fading Parameters}
The integral expressions in \eqref{eq:sep_mpsk_fas} and \eqref{eq:sep_mqam_fas} cannot be simplified further in closed-form without assuming integer multiplicities of the eigenvalues. To this end, to study in more detail the joint impact of the various system and channel parameters on the SEP performance, we further focus on the case of Nakagami-$m$ channels with identical fading parameters, i.e., $m_\ell=m$, $\forall \ell$. For this case, we have $K=L$, $\tilde{K}=\tilde{L}$, and each $[\boldsymbol{\lambda}]_q$ has a multiplicity of $2 \hat{m}$, implying $[\boldsymbol{\mu}]_q = 2 \hat{m}$, $\forall q$. This results in the expressions for the SEP with $M$-PSK and $M$-QAM given by
\begin{align}
&P_{e,{\rm FAS}}^{{\rm PSK},\hat{m}} \triangleq \frac{1}{\pi} \int_0^{\frac{\pi (M-1)}{M}} \! \!
\prod_{\ell=1}^{\tilde{L}} \left( \frac{\sin^2 (\zeta)}
{\sin^2 (\zeta) + \eta_{\rm PSK} \left( \overline{\gamma} \right) [\boldsymbol{\lambda}]_\ell} \right)^{\hat{m}} \nn \\
& \times \! \! \! \prod_{\ell=\tilde{L}+1}^{L} \! \! \left( \frac{\ell \sin^2 (\zeta)}
{\ell \sin^2 (\zeta) + \tilde{L} \eta_{\rm PSK} \left( \overline{\gamma} \right) [\boldsymbol{\lambda}]_\ell} \right)^{\hat{m}} \! \! {\rm d} \zeta
\label{eq:sep_mpsk_fas_same_m_1}
\end{align}
and
\begin{align}
& P_{e, {\rm FAS}}^{{\rm QAM},\hat{m}}\triangleq \frac{1}{2\pi M} \nn \\
& \times \! \! \left( \! \sum_{i=1}^{d_M} \omega_i^{\rm a} \! \! \int\limits_0^{\pi/2} \!
\prod_{\ell=1}^{\tilde{L}} \! \! \left( \! \frac{\sin^2 \left( \zeta \! + \! \frac{\pi}{4} \right)}
{\sin^2 \left(\zeta \! + \! \frac{\pi}{4} \right) \! + \! \eta_{i, {\rm QAM}} \left(\overline{\gamma}\right)[\boldsymbol{\lambda}]_\ell} \! \right)^{\! \! \hat{m}} \right. \nn \\
& \times \! \! \! \prod_{\ell=\tilde{L}+1}^{L} \! \! \left( \frac{\ell \sin^2 \left( \zeta \! + \! \frac{\pi}{4} \right)}
{\ell \sin^2 \left( \zeta \! + \! \frac{\pi}{4} \right) \! + \! \tilde{L} \eta_{i,\rm QAM} \left( \overline{\gamma} \right) [\boldsymbol{\lambda}]_\ell} \right)^{\! \! \hat{m}} \! \! {\rm d} \zeta \nn \\
& + \omega_i^{\rm b} \! \! \! \int\limits_0^{3\pi/4}
\prod_{\ell=1}^{\tilde{L}} \left(\frac{\sin^2 (\zeta)}
{\sin^2 (\zeta) + \eta_{i,{\rm QAM}} \left(\overline{\gamma}\right)[\boldsymbol{\lambda}]_\ell}\right)^{\! \! \hat{m}} \nn \\
& \left. \times \! \! \!
\prod_{\ell=\tilde{L}+1}^{L} \! \! \left( \! \frac{\ell \sin^2 (\zeta)}
{\ell \sin^2 \! (\zeta) \! + \! \tilde{L} \eta_{i,\rm QAM} \left( \overline{\gamma} \right) [\boldsymbol{\lambda}]_\ell} \! \right)^{\! \! \hat{m}} \! \! {\rm d} \zeta \right) ,
\label{eq:sep_mqam_fas_same_m_1}
\end{align}
respectively.

To further simplify these integral-form SEP expressions, we introduce the following definitions:
\begin{align*}
\boldsymbol{\Phi}^1_{\rm PSK} &\dn \left[ \Phi^1_{{\rm PSK},1}, \Phi^1_{{\rm PSK},2}, \ldots, \Phi^1_{{\rm PSK},\tilde{L}} \right]^{\mathrm{T}},\\
\boldsymbol{\Phi}^2_{\rm PSK} &\dn \left[ \Phi^1_{{\rm PSK},\tilde{L}+1}, \Phi^1_{{\rm PSK},\tilde{L}+2}, \ldots, \Phi^1_{{\rm PSK},L} \right]^{\mathrm{T}},\\
\boldsymbol{\Phi}^1_{i,\rm QAM} &\dn \left[ \Phi^1_{i,{\rm QAM},1}, \Phi^1_{i,{\rm QAM},2}, \ldots, \Phi^1_{i,{\rm QAM},\tilde{L}} \right]^{\mathrm{T}}, \\
\boldsymbol{\Phi}^2_{i,\rm QAM} &\dn \left[ \Phi^2_{i,{\rm QAM},\tilde{L}+1}, \Phi^2_{i,{\rm QAM},\tilde{L}+2}, \ldots, \Phi^2_{i,{\rm QAM},L} \right]^{\mathrm{T}},
\end{align*}
with $\Phi^1_{{\rm PSK},\ell} \dn \left(\eta_{\rm PSK} \left( \overline{\gamma} \right) [\boldsymbol{\lambda}]_\ell \right)^{-1}$, $\forall \ell=1,\ldots, \tilde{L}$, $\Phi^2_{{\rm PSK},\ell} \dn \big( \frac{\tilde{L}}{\ell} \eta_{\rm PSK} \left( \overline{\gamma} \right) [\boldsymbol{\lambda}]_\ell \big)^{-1}$, $\forall \ell=\tilde{L}+1,\ldots,L$, $\Phi^1_{i,{\rm QAM},\ell} \dn \left( \eta_{i, {\rm QAM}} \left(\overline{\gamma}\right)[\boldsymbol{\lambda}]_\ell \right)^{-1}$, $\forall i,\ell=1,\ldots, \tilde{L}$, and $\Phi^2_{i,{\rm QAM},\ell} \dn \big( \frac{\tilde{L}}{\ell} \eta_{i, {\rm QAM}} \left(\overline{\gamma}\right)[\boldsymbol{\lambda}]_\ell \big)^{-1}$, $\forall i,\ell=\tilde{L}+1,\ldots, L$. Using these definitions,~\eqref{eq:sep_mpsk_fas_same_m_1} and~\eqref{eq:sep_mqam_fas_same_m_1} can be rewritten as
\begin{align}
P_{e,{\rm FAS}}^{{\rm PSK},\hat{m}} = \ & \frac{1}{\pi} \int_0^{\frac{\pi (M-1)}{M}} \! \!
\prod_{\ell=1}^{\tilde{L}} \left( \frac{\Phi^1_{{\rm PSK},\ell}}
{\Phi^1_{{\rm PSK},\ell} + \sin^{-2} (\zeta)} \right)^{\hat{m}} \nn \\
& \times \! \! \! \prod_{\ell=\tilde{L}+1}^{L} \! \! \left( \frac{\Phi^2_{{\rm PSK},\ell}}
{\Phi^2_{{\rm PSK},\ell} + \sin^2 (\zeta)} \right)^{\hat{m}} \! \! {\rm d} \zeta,
\label{eq:sep_mpsk_fas_same_m}
\end{align}
and
\begin{align}
& P_{e, {\rm FAS}}^{{\rm QAM},\hat{m}}= \frac{1}{2\pi M}  \nn \\
& \times \! \! \left( \sum_{i=1}^{d_M} \omega_i^{\rm a} \int\limits_0^{\pi/2}
\prod_{\ell=1}^{\tilde{L}} \left( \frac{\Phi^1_{i,{\rm QAM},\ell}}
{\Phi^1_{i,{\rm QAM},\ell} + \sin^{-2} \left(\zeta + \frac{\pi}{4} \right)} \right)^{\hat{m}} \right. \nn \\
& \times \prod_{\ell=\tilde{L}+1}^{L} \! \! \left( \frac{\Phi^2_{i,{\rm QAM},\ell}}
{\Phi^2_{i,{\rm QAM},\ell} + \sin^{-2} \left( \zeta + \frac{\pi}{4} \right)} \right)^{\hat{m}} {\rm d} \zeta \nn \\
& + \omega_i^{\rm b} \int\limits_0^{3\pi/4}
\prod_{\ell=1}^{\tilde{L}} \left(\frac{\Phi^1_{i,{\rm QAM},\ell}}
{\Phi^1_{i,{\rm QAM},\ell} + \sin^2 (\zeta)}\right)^{\hat{m}} \nn \\
& \left. \times
\prod_{\ell=\tilde{L}+1}^{L} \left( \frac{\Phi^2_{i,{\rm QAM},\ell}}
{\Phi^2_{i,{\rm QAM},\ell} + \sin^{-2} (\zeta)} \right)^{\hat{m}} {\rm d} \zeta \right),
\label{eq:sep_mqam_fas_same_m}
\end{align}
respectively. These can be simplified further by making use of the following proposition, including the decomposition method of products of functions as appearing in \eqref{eq:sep_mpsk_fas_same_m} and \eqref{eq:sep_mqam_fas_same_m}.

\begin{proposition}
\label{prop:decomposition_method}
Consider the function
\beq
f(x) \dn \prod_{\ell=1}^L \left( \frac{a_\ell}{a_\ell + x} \right)^{\hat{m}} ,
\label{eq:funcx}
\eeq
with $\{a_\ell\}$ representing its $L$ distinct poles each with algebraic multiplicity of $\hat{m}$. Using the definition $\undb{a} \dn [a_1,a_2,\ldots,a_L]^{\mathrm{T}}$, $f(x)$ can be decomposed as
\beq
f(x) = \sum_{\ell=1}^L \sum_{k=1}^{\hat{m}} \alpha_{\ell,k} (\undb{a},\hat{m},L)
\left( \frac{a_\ell}{a_\ell + x} \right)^{k},
\label{eq:decomp_funcx}
\eeq
with the coefficients given by
\begin{align}
& \alpha_{\ell,k} (\undb{a},\hat{m},L) \dn \left( \prod_{\substack{p=1\\ p\neq \ell}}^{L}
\left(\frac{a_p}{a_p-a_\ell}\right)^{\hat{m}} \right) \label{eq:alpha_coeff} \nn \\
& \times \! \! \! \! \! \! \! \!
\sum_{\substack{(q_1,\ldots,q_{\hat{m}-k}) \\ 0 \leq q_1,\ldots,q_{\hat{m}-k} \leq \hat{m}-k \\
q_1 + 2q_2 + \ldots + (\hat{m}-k) q_{\hat{m}-k}=\hat{m}-k}}
\! \! \! \! \! \! \! \! \! \! \! \! \prod_{r=1}^{\hat{m}-k} \frac{1}{q_r!}
\left( \frac{\hat{m}}{r} \sum_{\substack{p=1 \\ p \neq \ell}}^L
\left( \frac{a_\ell}{a_\ell-a_p} \right) \right)^{q_r} \! \! .
\end{align}
The composite summation in this expression is carried out for all tuples of $(q_1,\ldots,q_{\hat{m}-k})$ satisfying $0 \leq q_1,\ldots,q_{\hat{m}-k} \leq \hat{m}-k$ and $q_1 + 2q_2 + \ldots + (\hat{m}-k) q_{\hat{m}-k}=\hat{m}-k$.
\end{proposition}

\begin{IEEEproof}
Employing the Heaviside theorem for partial fraction expansion, $f(x)$ can be expressed as the weighted sum of elementary functions of the form $(a_\ell/(a_\ell + x))^k$, which can be expressed using the decomposition formula \cite{1025509, kreyszig1998advanced} as in~\eqref{eq:decomp_funcx}. Furthermore, the coefficient $\alpha_{\ell,k}$, $\forall \ell=1,\ldots,L$ and $\forall k=1,\ldots,\hat{m}$, is obtained as
\beq
\! \! \! \alpha_{\ell,k} (\undb{a},\hat{m},L) = \frac{1}{a_{\ell}^k (\hat{m}-k)!}
\! \! \left. \frac{{\rm d}^{\hat{m}-k} \{x^{\hat{m}} f(x-a_\ell)\}}
{{\rm d} x^{\hat{m}-k}} \right|_{x=0} .
\label{eq:alpha_coeff_step}
\eeq
The proof is completed by using Faa di Bruno's formula~\cite{poussin1959cours} to compute the derivatives in \eqref{eq:alpha_coeff_step}, followed by algebraic simplifications.
\end{IEEEproof}

Using Proposition~\ref{prop:decomposition_method}, we present the following proposition to compute the integrals involving the elementary functions in \eqref{eq:sep_mpsk_fas_same_m} and \eqref{eq:sep_mqam_fas_same_m}.
\begin{proposition}
\label{prop:sep_integral}
The integral $I(\Theta) \triangleq \frac{1}{\pi}\int_0^{\Theta} f(\sin^{-1}(\zeta)) {\rm d} \zeta$, with $f(x)$ as in~\eqref{eq:funcx}, can be computed using Proposition~\ref{prop:decomposition_method} as
\begin{align}
I (\Theta) = \sum_{\ell=1}^L \sum_{k=1}^{\hat{m}} \alpha_{\ell,k} (\undb{a},\hat{m},L)
{\mathcal{F}}_{\ell,k} (a_{\ell}, \Theta),
\label{eq:sep_int}
\end{align}
with
\begin{align}
{\mathcal{F}}_{\ell,k} (a_{\ell}, \Theta) & \dn \frac{1}{\pi} \int_0^{\Theta}
\left( \frac{\sin^2 (\zeta)}{a_{\ell}^{-1} + \sin^2 (\zeta)} \right)^{k} {\rm d} \zeta \nn \\
& = \frac{\Theta}{\pi} + \sum_{j=1}^k (-1)^j \binom{k}{j} {\mathcal{G}}_{\ell,j} (a_\ell, \Theta) ,
\label{eq:ffcun_int}
\end{align}
where each function ${\mathcal{G}}_{\ell,j} (a_\ell, \Theta)$ is given by
\begin{align}
& {\mathcal{G}}_{\ell,j} (a_\ell, \Theta) \dn \frac{\pi^{-1}}{ (1+a_\ell)^{j-1/2}}
\sum_{q=0}^{j-1} \binom{j-1}{q} \binom{2q}{q} \left( \frac{a_\ell}{4} \right)^q \nn \\
& \times \left( \pi - \tan^{-1} \left( \sqrt{1+a_\ell} |\tan(\Theta)| \right) 
+ \sqrt{1+a_\ell} \frac{\tan(\Theta)}{2} \right. \nn \\
& \left. \times \sum_{p=1}^q \frac{4^p}{\binom{2p}{p} p \left(1 + (1+a_\ell) \tan^2(\Theta) \right)^p} \right).
\label{eq:gfunc_int}
\end{align}
\end{proposition}

\begin{IEEEproof}
By using the binomial theorem expansion in the integral form inside~\eqref{eq:ffcun_int}, this expression can be rewritten as
\beq
{\mathcal{F}}_{\ell,k} (a_{\ell}, \zeta)
\!=\! \frac{\zeta}{\pi} \! +\!  \sum_{j=1}^k (-1)^j \binom{k}{j} \underbrace{\int \frac{\pi^{-1}}{\left( 1\! +\!  a_\ell \sin^2 (\zeta) \right)^j} {\rm d} \zeta}_{={\mathcal{G}}_{\ell,j} (a_\ell, \zeta)}.
\label{eq:ffcun_int_indefinite2}
\eeq
Making the change of variable $\tan (\zeta) = \frac{\tan (\phi)}{\sqrt{1+a_\ell}}$ in the latter integral and performing some algebraic simplifications yields 
\beq
{\mathcal{G}}_{\ell,j} (a_\ell, g(\phi)) = \frac{\pi^{-1}}{ (1+a_\ell)^{j-1/2}}
\sum_{q=0}^{j-1} \binom{j-1}{q} a_{\ell}^q {\mathcal{K}}_{2q} (\phi) ,
\label{eq:gfunc_int_indefinite2}
\eeq
with $g(\phi) \dn \tan^{-1} \left( \frac{\tan (\phi)}{\sqrt{1+a_\ell}} \right)=\zeta$ and where the integral ${\mathcal{K}}_{2q} (\phi) \dn \int \cos^{2q} (\phi){\rm d} \phi$ is solved using \cite[eq. 2.512(2)]{gradshteyn2000table} as
\begin{align}
{\mathcal{K}}_{2q} (\phi) & = \frac{(2q)!}{4^q (q!)^2} \left( \phi + \frac{\sin (\phi)}{2}
\sum_{p=1}^q \frac{4^p (p!)^2 \cos^{2p-1} (\phi)}{p (2p)!} \right) \nn \\
& = \frac{\binom{2q}{q}}{4^q} \left( \phi + \frac{\tan (\phi)}{2}
\sum_{p=1}^q \frac{4^p}{\binom{2p}{p} p \left( 1+ \tan^2 (\phi) \right)^p} \right) .
\label{eq:kfunc}
\end{align}
Performing the substitution $\phi = g^{-1} (\zeta)$ in~\eqref{eq:gfunc_int_indefinite2} followed by algebraic simplifications gives ${\mathcal{G}}_{\ell,j} (a_\ell, \zeta)$, which yields~\eqref{eq:gfunc_int} when evaluated at $\zeta=0$, completing the proof.
\end{IEEEproof}

By applying Proposition~\ref{prop:decomposition_method} to the SEP formula in~\eqref{eq:sep_mpsk_fas_same_m} valid for $M$-PSK, we obtain
\begin{align}
&
P_{e,{\rm FAS}}^{{\rm PSK}, \hat{m}} = \frac{1}{\pi} \! \! \! \! \! \int\limits_0^{\frac{\pi(M-1)}{M}} \! \! \! \!
\sum_{\ell_1=1}^{\tilde{L}} \sum_{\ell_2=\tilde{L}+1}^{L} \sum_{k_1=1}^{\hat{m}} \sum_{k_2=1}^{\hat{m}}
\! \alpha_{\ell_1,k_1} \big( \boldsymbol{\Phi}^1_{\rm PSK} , \hat{m}, \tilde{L} \big) \nn \\
& \times \alpha_{\ell_2,k_2} \big( \boldsymbol{\Phi}^2_{\rm PSK} , \hat{m}, L-\tilde{L} \big)
\left( \frac{\Phi^1_{{\rm PSK},\ell_1}}{\Phi^1_{{\rm PSK},\ell_1} + \sin^{-2} (\zeta)} \right)^{k_1} \nn \\
& \times \left( \frac{\Phi^2_{{\rm PSK},\ell_2}}
{\Phi^2_{{\rm PSK},\ell_2} + \sin^{-2} (\zeta)} \right)^{k_2} {\rm d} \zeta .
\label{eq:sep_mpsk_fas_same_m2}
\end{align}

\begin{proposition}
\label{prop:sin_integration}
The integral ${\mathcal{J}} (\Theta, \hat{m}, L) \dn \int_0^\Theta \sin^{2 \hat{m} L} (\zeta){\rm d} \zeta$
is solved for integer $\hat{m}$ and $L$ as
\begin{align}
{\mathcal{J}} (\Theta, \hat{m}, L) = \ & \frac{\Theta}{2^{2 \hat{m} L}}\binom{2 \hat{m} L}{\hat{m} L} \nn\\
& + \frac{1}{2^{2 \hat{m} L}} \sum_{k=1}^{\hat{m} L} \frac{(-1)^k}{k}
\binom{2 \hat{m} L}{\hat{m} L-k} \sin (2 k \Theta) . \label{eq:ifunc_val}
\end{align}
\end{proposition}

\begin{IEEEproof}
The function $\sin^{2 \hat{m} L}(\zeta)$ appearing in ${\mathcal{J}} (\Theta, \hat{m}, L)$ can be expanded for integer $\hat{m}$ and $L$ as
\begin{align} 
\sin^{2 \hat{m} L} \zeta \! = \frac{\binom{2 \hat{m} L}{\hat{m} L}}{2^{2 \hat{m} L}}
\! + \! \frac{1}{2^{2 \hat{m} L - 1}} \sum_{k=1}^{\hat{m} L} (-1)^k
\binom{2 \hat{m} L}{\hat{m} L-k} \! \cos (2 k \zeta) .
\end{align}
Replacing with this expansion and applying some algebraic simplifications completes the proof.
\end{IEEEproof}

By re-applying Proposition~\ref{prop:decomposition_method} as well as Proposition~\ref{prop:sin_integration} in \eqref{eq:sep_mpsk_fas_same_m2}, we obtain the closed-form expression for the SEP with $M$-PSK of a GSC$(\tilde{L},L)$-based FAS receiver as
\begin{align}
P_{e,{\rm FAS}}^{{\rm PSK},\hat{m}} = \ & 
\sum_{\ell_1=1}^{\tilde{L}} \sum_{k_1=1}^{\hat{m}}
\alpha_{\ell_1,k_1} \big( \boldsymbol{\Phi}^1_{\rm PSK} , \hat{m}, \tilde{L} \big) \nn \\
& \times \sum_{\ell_2=\tilde{L}+1}^{L} \sum_{k_2=1}^{\hat{m}}
\alpha_{\ell_2,k_2} \big( \boldsymbol{\Phi}^2_{\rm PSK} , \hat{m}, L-\tilde{L} \big) \nn \\
&  \times \sum_{n=1}^2 \sum_{\ell_3=1}^{k_n}
\alpha_{\ell_2,k_2} \big( \boldsymbol{\Phi}^{\ell_1,\ell_2}_{\rm PSK} , \hat{m}, 2 \big) \nn \\
& \times {\mathcal{F}}_{n,\ell_3} \left( \big[ \boldsymbol{\Phi}_{{\rm PSK}}^{\ell_1,\ell_2} \big]_n , \frac{\pi(M-1)}{M} \right) ,
\label{eq:sep_mpsk_fas_same_m_final}
\end{align}
with $\boldsymbol{\Phi}^{\ell_1,\ell_2}_{\rm PSK} \dn \big[ \Phi_{{\rm PSK},\ell_1}^{1}, \Phi_{{\rm PSK},\ell_2}^{2} \big]^{\mathrm{T}}$. 

Similarly, starting from \eqref{eq:sep_mqam_fas_same_m}, the SEP with $M$-QAM of a GSC$(\tilde{L},L)$-based FAS receiver is obtained as 
\begin{align}
P_{e,{\rm FAS}}^{{\rm QAM},\hat{m}} = \ & \frac{1}{2M} \sum_{i=1}^{d_M}
\sum_{\ell_1=1}^{\tilde{L}} \sum_{k_1=1}^{\hat{m}}
\alpha_{\ell_1,k_1} \big( \boldsymbol{\Phi}^1_{i, \rm QAM} , \hat{m}, \tilde{L} \big) \nn \\
&  \times \sum_{\ell_2=\tilde{L}+1}^{L} \sum_{k_2=1}^{\hat{m}}
\alpha_{\ell_2,k_2} \big( \boldsymbol{\Phi}^2_{i, \rm QAM} , \hat{m}, L-\tilde{L} \big) \nn \\
&  \times \sum_{n=1}^2 \sum_{\ell_3=1}^{k_n}
\alpha_{\ell_2,k_2} \big( \boldsymbol{\Phi}^{\ell_1,\ell_2}_{i, \rm QAM} , \hat{m}, 2 \big) \nn \\
& \times \left( \omega_i^{\rm a}
{\tilde{\mathcal{F}}}_{n,\ell_3} \left( \big[ \boldsymbol{\Phi}_{{i,\rm QAM}}^{\ell_1,\ell_2} \big]_n , \frac{\pi}{2} \right) \right. \nn \\
& \left. + \omega_i^{\rm b} {\mathcal{F}}_{n,\ell_3} \left( \big[ \boldsymbol{\Phi}_{{i,\rm QAM}}^{\ell_1,\ell_2} \big]_n , \frac{3\pi}{4} \right) \right) ,
\label{eq:sep_mqam_fas_same_m2}
\end{align}
with $\boldsymbol{\Phi}^{\ell_1,\ell_2}_{i, \rm QAM} \dn \big[ \Phi_{i,{\rm QAM},\ell_1}^{1}, \Phi_{i,{\rm QAM},\ell_2}^{2} \big]^{\mathrm{T}}$ and
\beq
\tilde{{\mathcal{F}}}_{\ell,k} \left( a, \frac{\pi}{2} \right)
\dn \frac{1}{\pi} \int_0^{\frac{\pi}{2}}
\left(\frac{\sin^2 \left( \zeta + \pi/4 \right)}{\sin^2 \left(\zeta+ \pi/4 \right)
+ a^{-1}} \right)^{k} {\rm d} \zeta .
\label{eq:ftilde_func}
\eeq
Employing the change of variable $\tilde{\zeta}\dn\zeta+\pi/4$ yields
\begin{align}
& \tilde{{\mathcal{F}}}_{\ell,k} \left( a, \frac{\pi}{2} \right)
= \frac{1}{\pi} \int_{\frac{\pi}{4}}^{\frac{3\pi}{4}} \! \!
\left(\frac{\sin^2 (\tilde{\zeta})}{\sin^2 (\tilde{\zeta})
+ a^{-1}} \right)^{k} \! {\rm d} \zeta \nn \\
& = \frac{1}{\pi} \! \left( \int\limits_0^{\frac{3 \pi}{4}}
\! \! \left(\frac{\sin^2 (\tilde{\zeta})}{\sin^2 (\tilde{\zeta})
+ a^{-1}} \right)^{\!k} \! \! {\rm d} \zeta 
\!-\! \int\limits_0^{\frac{\pi}{4}} \left(\frac{\sin^2 (\tilde{\zeta})}{\sin^2 (\tilde{\zeta})
+ a^{-1}} \right)^{\!k} \! \! {\rm d} \zeta \right) \nn \\
& = {\mathcal{F}}_{\ell,k} \left( a, \frac{3\pi}{4} \right)
- {\mathcal{F}}_{\ell,k} \left( a, \frac{\pi}{4} \right),
\label{eq:ftilde_func2}
\end{align}
which can be substituted into \eqref{eq:sep_mqam_fas_same_m2} to provide the closed-form expression as
\begin{align}
P_{e,{\rm FAS}}^{{\rm QAM},\hat{m}} = \ & \frac{1}{2M} \sum_{i=1}^{d_M}
\sum_{\ell_1=1}^{\tilde{L}} \sum_{k_1=1}^{\hat{m}}
\alpha_{\ell_1,k_1} \big( \boldsymbol{\Phi}^1_{i, \rm QAM} , \hat{m}, \tilde{L} \big) \nn \\
&  \times \sum_{\ell_2=\tilde{L}+1}^{L} \sum_{k_2=1}^{\hat{m}}
\alpha_{\ell_2,k_2} \big( \boldsymbol{\Phi}^2_{i, \rm QAM} , \hat{m}, L-\tilde{L} \big) \nn \\
&  \times \sum_{n=1}^2 \sum_{\ell_3=1}^{k_n}
\alpha_{\ell_2,k_2} \big( \boldsymbol{\Phi}^{\ell_1,\ell_2}_{i, \rm QAM} , \hat{m}, 2 \big) \nn \\
&  \times \left( \left( \omega_i^{\rm a} + \omega_i^{\rm b} \right)
{\mathcal{F}}_{n,\ell_3} \left( \big[ \boldsymbol{\Phi}_{{i,\rm QAM}}^{\ell_1,\ell_2} \big]_n , \frac{3\pi}{4} \right) \right. \nn \\
& \left. - \omega_i^{\rm a} {\mathcal{F}}_{n,\ell_3} \left( \big[ \boldsymbol{\Phi}_{{i,\rm QAM}}^{\ell_1,\ell_2} \big]_n , \frac{\pi}{4} \right) \right) .
\label{eq:sep_mqam_fas_same_m_final}
\end{align}
\subsection{Asymptotic Expressions}
In the regime of high average SNR, i.e., for $\overline{\gamma} \gg 1$, we have $\eta_{\rm PSK} (\overline{\gamma})[\boldsymbol{\lambda}]_\ell \gg \sin^2 (\zeta)$ and $\eta_{i, {\rm QAM}} (\overline{\gamma})[\boldsymbol{\lambda}]_\ell \gg \sin^2 (\zeta) , \, \sin^2 (\zeta + \pi/4)$, $\forall \ell=1,2,\ldots,L$. This results in the asymptotic version of the SEP expression in~\eqref{eq:sep_mpsk_fas_same_m_final} for $M$-PSK given by
\begin{align}
& P_{e,{\rm FAS} \big| \overline{\gamma} \gg 1}^{{\rm PSK},\hat{m}}
= \frac{1}{\pi} \int_0^{\frac{\pi (M-1)}{M}} \! \!
\prod_{\ell=1}^{\tilde{L}} \left( \frac{\sin^2 (\zeta)}
{\eta_{\rm PSK} \left( \overline{\gamma} \right) [\boldsymbol{\lambda}]_\ell} \right)^{\hat{m}} \nn \\
& \times \! \! \!
\prod_{\ell=\tilde{L}+1}^{L} \! \! \left( \frac{\ell \sin^2 (\zeta)}
{\tilde{L} \eta_{\rm PSK} \left( \overline{\gamma} \right) [\boldsymbol{\lambda}]_\ell} \right)^{\hat{m}} \! \! {\rm d} \zeta \nn \\
& = \frac{1}{\pi} \int_0^{\frac{\pi (M-1)}{M}} \! \! \! \!
\frac{\big( \frac{L!}{\tilde{L}!}\big)^{\hat{m}} \sin^{2 \hat{m} L} (\zeta)}
{\tilde{L}^{\hat{m} (L-\tilde{L})} \eta_{\rm PSK}^{\hat{m} L} (\overline{\gamma})
\left(\prod_{\ell=1}^L[\boldsymbol{\lambda}]^{\hat{m}}_\ell\right)} {\rm d} \zeta .
\label{eq:sep_fas_mpsk_high_snr1}
\end{align}
Using Proposition~\ref{prop:sin_integration}, the latter expression further simplifies to
\begin{align}
P_{e,{\rm FAS} \big| \overline{\gamma} \gg 1}^{{\rm PSK},\hat{m}} = \ & \frac{\big( \frac{L!}{\tilde{L}!} \big)^{\hat{m}} \left( N_{\text{p}} \varepsilon + 1\right)^{\hat{m} L}
{\mathcal{J}} \big( \frac{\pi(M-1)}{M}, \hat{m}, L \big)}
{\pi \overline{\gamma}^{\hat{m} L} \tilde{L}^{\hat{m} (L-\tilde{L})}
\left( c_{\rm PSK} N_{\text{p}} \varepsilon \right)^{\hat{m} L}} \nonumber \\
&\times\left(\prod_{\ell=1}^L[\boldsymbol{\lambda}]^{-\hat{m}}_\ell\right).
\label{eq:sep_fas_mpsk_high_snr}
\end{align}
Following similar steps for the SEP expression in~\eqref{eq:sep_mqam_fas_same_m_final} for $M$-QAM and using the derivation $\int_0^{\frac{\pi}{2}} \! \! \! \sin^{2 \hat{m} L} \left( \zeta + \frac{\pi}{4} \right) {\rm d} \zeta$ $=$$\int_{\frac{\pi}{4}}^{\frac{3\pi}{4}} \! \! \! \sin^{2 \hat{m} L} (\zeta) {\rm d} \zeta$$=$$\int_0^{\frac{3\pi}{4}} \! \! \! \sin^{2 \hat{m} L} (\zeta) {\rm d} \zeta - \int_0^{\frac{\pi}{4}} \! \! \sin^{2 \hat{m} L} (\zeta) {\rm d} \zeta$, we obtain
\begin{align}
& \! \! \! P_{e,{\rm FAS} \big| \overline{\gamma} \gg 1}^{{\rm QAM},\hat{m}} \approx \frac{\big( \frac{L!}{\tilde{L}!} \big)^{\hat{m}}
\left(\prod_{\ell=1}^L[\boldsymbol{\lambda}]^{-\hat{m}}_\ell\right)}
{2 \pi M \overline{\gamma}^{\hat{m} L} \tilde{L}^{\hat{m} (L-\tilde{L})} c_{\rm QAM}^{\hat{m} L}}
\sum_{i=1}^{d_M} \left( \frac{N_{\text{p}} \varepsilon + \psi_i}{N_{\text{p}} \varepsilon} \right)^{\hat{m} L} \nn \\
& \! \! \! \times \left( \left( \omega_i^{\rm a} + \omega_i^{\rm b} \right) 
{\mathcal{J}} \left( \frac{3 \pi}{4}, \hat{m}, L \right)
- \omega_i^{\rm b} {\mathcal{J}} \left( \frac{\pi}{4}, \hat{m}, L \right) \right) .
\label{eq:sep_fas_mqam_high_snr}
\end{align}

\begin{figure*}[!t]
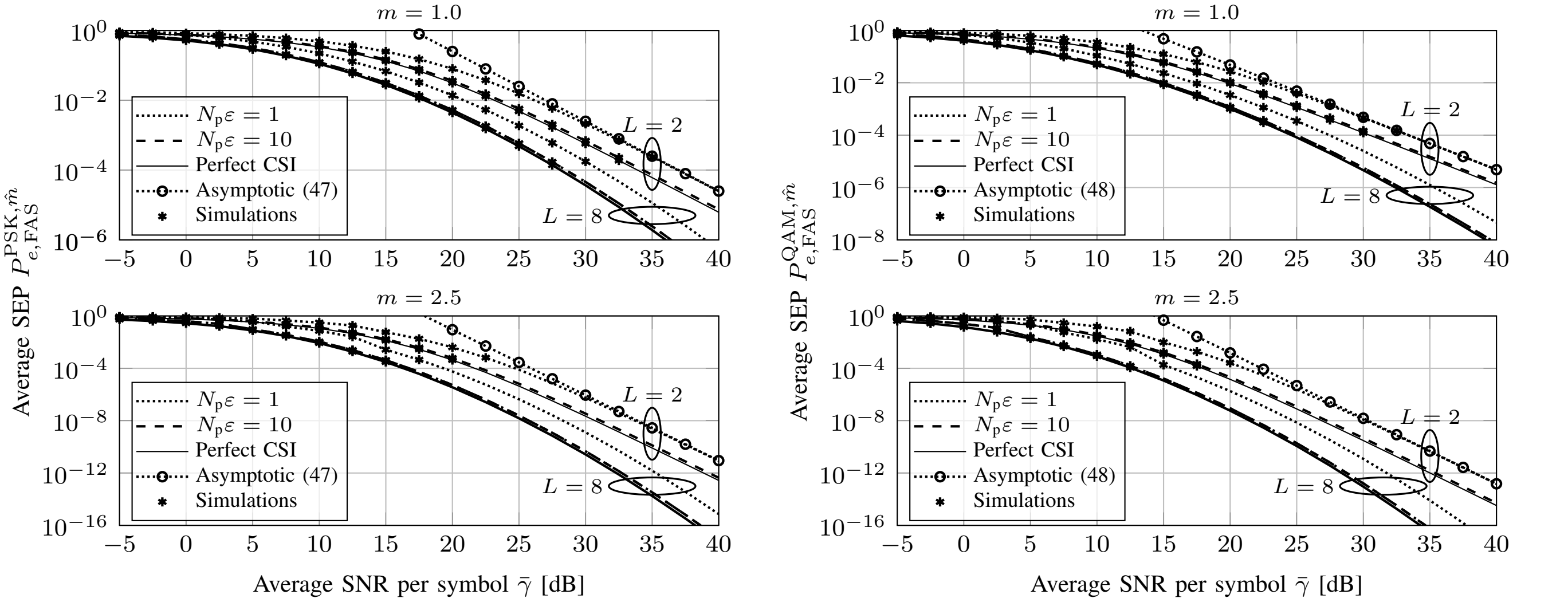

\begin{subfigure}[b]{\columnwidth}
\centering
\input{Figures/plot4.tex}\vspace{-2mm}
\caption{$16$-PSK.}
\label{f2c}
\end{subfigure}
\begin{subfigure}[b]{\columnwidth}
\centering
\input{Figures/plot5.tex}\vspace{-2mm}
\caption{$16$-QAM.}
\label{f2d}
\end{subfigure}
\caption{Average SEP with (a) $16$-PSK and (b) $16$-QAM of a GSC$(1,L)$-based FAS receiver versus average SNR per symbol, with $W=0.2$ and different values of $m$, $L$, and $N_{\text{p}}\varepsilon$.}
\vspace{-3mm}
\label{f2}
\end{figure*}
Additionally, for a FAS receiver with a large number of ports (i.e., large $L$), the following approximation holds for $M>2$:
\begin{align}
{\mathcal{J}} \left( \frac{\pi (M-1)}{M}, \hat{m}, L \right) & \approx \sqrt{\frac{\pi}{\hat{m} L}}, \nn \\
{\mathcal{J}} \left( \frac{\pi}{4}, \hat{m}, L \right) & \approx \frac{2^{-\hat{m}L}}{2 \hat{m} L}.
\label{eq:largeL_approx}
\end{align}
For this case, the asymptotic expressions for the SEP with $M$-PSK and $M$-QAM can be obtained as
\begin{align}
& P_{e,{\rm FAS} \big| \overline{\gamma},L \gg 1}^{{\rm PSK},\hat{m}}
\approx \ \frac{\big( \frac{L!}{\tilde{L}!} \big)^{\hat{m}} (N_{\text{p}} \varepsilon + 1)^{\hat{m} L} \big(\prod_{\ell=1}^L[\boldsymbol{\lambda}]^{-\hat{m}}_\ell \! \big)}
{\sqrt{\pi \hat{m} L} \tilde{L}^{\hat{m} (L-\tilde{L})}
(c_{\rm PSK} N_{\text{p}} \varepsilon)^{\hat{m} L} \overline{\gamma}^{\hat{m} L}}  ,
\label{eq:sep_fas_mpsk_high_snr_L} \\
& P_{e,{\rm FAS} \big| \overline{\gamma} \gg 1}^{{\rm QAM},\hat{m}}
\! \approx \ \frac{\big( \frac{L!}{\tilde{L}!} \big)^{\hat{m}}
\big(\prod_{\ell=1}^L[\boldsymbol{\lambda}]^{-\hat{m}}_\ell\big)}
{2 \pi M \overline{\gamma}^{\hat{m} L} \tilde{L}^{\hat{m} (L-\tilde{L})} c_{\rm QAM}^{\hat{m} L}}
\sum_{i=1}^{d_M} \! \left( \! \frac{N_{\text{p}} \varepsilon + \psi_i}{N_{\text{p}} \varepsilon} \! \right)^{\! \! \hat{m} L} \nn \\
& \qquad \qquad \qquad
\times \left( \left( \omega_i^{\rm a} + \omega_i^{\rm b} \right) \sqrt{\frac{\pi}{\hat{m} L}}
- \frac{2^{-\hat{m}L} \omega_i^{\rm b}}{2 \hat{m} L} \right) ,
\label{eq:sep_fas_mqam_high_snr_L}
\end{align}
respectively.

From \eqref{eq:sep_fas_mpsk_high_snr}--\eqref{eq:sep_fas_mqam_high_snr} and \eqref{eq:sep_fas_mpsk_high_snr_L}--\eqref{eq:sep_fas_mqam_high_snr_L}, it can be concluded that imperfect CSI for the considered GSC$(\tilde{L},L)$-based FAS receiver degrades the coding gain, whereas the diversity gain depends only on $L$ and $\hat{m}$.
\begin{figure}[!t]
\centering
\pgfdeclarelayer{background}
\pgfdeclarelayer{foreground}
\pgfsetlayers{background,main,foreground}

\begin{tikzpicture}

\begin{groupplot}[
    group style={
        group size=1 by 2,
        vertical sep=25pt,
        y descriptions at=edge left,
    },
    width=8.5cm,
    height=5cm,
    xlabel near ticks,
    ylabel near ticks,
    ylabel style={yshift=-3pt},
    label style={font=\footnotesize},
    ticklabel style={font=\footnotesize},
    legend style={
        at={(0.02,0.02)},
        anchor=south west,
        font=\scriptsize,
        inner sep=1pt,
        row sep=-2pt,
        fill opacity=0.75,
        draw opacity=1,
        text opacity=1
    },
    legend cell align=left,
    grid=major,
]
\nextgroupplot[
    xlabel={},
    ylabel={Average SEP $P_{e,{\rm FAS}}^{{\rm PSK},\hat{m}}$},
    xmin=1, xmax=10,
    xtick={1,2,...,10},
    ymin=1e-8, ymax=1, ymode=log,
    ytick={1,1e-2,1e-4,1e-6,1e-8},
    title={16-PSK},
    title style={font=\scriptsize,yshift=-2mm},
]
\addplot[black, thick, mark=asterisk, mark options={solid}, mark size=1.5pt]
table [x=mNak, y=Pe_mpsk_fas, col sep=comma] {Figures/files_txt/plot7/sep_m_mpsk_fas_16M_10gamdb_1Npe_4L_3Lhat_0.2W.txt};
\addlegendentry{$L = 4,\; W = 0.2$};

\addplot[black, thick, dashed, mark=o, mark options={solid}, mark size=1.5pt]
table [x=mNak, y=Pe_mpsk_fas, col sep=comma] {Figures/files_txt/plot7/sep_m_mpsk_fas_16M_10gamdb_1Npe_4L_3Lhat_0.8W.txt};
\addlegendentry{$L = 4,\; W = 0.8$};

\addplot[black, thick, densely dotted, mark=square, mark options={solid}, mark size=1.5pt]
table [x=mNak, y=Pe_mpsk_fas, col sep=comma] {Figures/files_txt/plot7/sep_m_mpsk_fas_16M_10gamdb_1Npe_8L_3Lhat_0.2W.txt};
\addlegendentry{$L = 8,\; W = 0.2$};

\addplot[black, thick, dash dot, mark=triangle, mark options={solid}, mark size=1.5pt]
table [x=mNak, y=Pe_mpsk_fas, col sep=comma] {Figures/files_txt/plot7/sep_m_mpsk_fas_16M_10gamdb_1Npe_8L_3Lhat_0.8W.txt};
\addlegendentry{$L = 8,\; W = 0.8$};
\addplot[black, thick, mark=asterisk, mark options={solid}, mark size=1.5pt]
table [x=mNak, y=Pe_mpsk_fas, col sep=comma] {Figures/files_txt/plot7/sep_m_mpsk_fas_16M_10gamdb_10Npe_4L_3Lhat_0.2W.txt};

\addplot[black, thick, dashed, mark=o, mark options={solid}, mark size=1.5pt]
table [x=mNak, y=Pe_mpsk_fas, col sep=comma] {Figures/files_txt/plot7/sep_m_mpsk_fas_16M_10gamdb_10Npe_4L_3Lhat_0.8W.txt};

\addplot[black, thick, densely dotted, mark=square, mark options={solid}, mark size=1.5pt]
table [x=mNak, y=Pe_mpsk_fas, col sep=comma] {Figures/files_txt/plot7/sep_m_mpsk_fas_16M_10gamdb_10Npe_8L_3Lhat_0.2W.txt};

\addplot[black, thick, dash dot, mark=triangle, mark options={solid}, mark size=1.5pt]
table [x=mNak, y=Pe_mpsk_fas, col sep=comma] {Figures/files_txt/plot7/sep_m_mpsk_fas_16M_10gamdb_10Npe_8L_3Lhat_0.8W.txt};
\node[draw, semithick, ellipse, minimum width=0.2cm, minimum height=0.8cm, inner sep=0pt] at (axis cs:9,1e-2) {};
\node[black, font=\scriptsize, anchor=center] at (axis cs:9,2.5e-1) {$N_{\text{p}} \varepsilon = 1$};
\node[draw, semithick, ellipse, minimum width=2cm, minimum height=0.2cm, inner sep=0pt, rotate=45] at (axis cs:7.5,1.5e-5) {};
\node[black,font=\scriptsize, anchor=center] at (axis cs:5.75,1e-7) {$N_{\text{p}} \varepsilon = 10$};
\nextgroupplot[
    xlabel={$m$-parameter},
    ylabel={Average SEP $P_{e,{\rm FAS}}^{{\rm QAM},\hat{m}}$},
    xmin=1, xmax=10,
    xtick={1,2,...,10},
    ymin=1e-12, ymax=1, ymode=log,
    ytick={1,1e-2,1e-4,1e-6,1e-8,1e-10,1e-12},
    title={16-QAM},
    title style={font=\scriptsize,yshift=-2mm},
]
\addplot[black, thick, mark=asterisk, mark options={solid}, mark size=1.5pt]
table [x=mNak, y=Pe_mqam_fas, col sep=comma] {Figures/files_txt/plot7/sep_m_mqam_fas_16M_10gamdb_1Npe_4L_3Lhat_0.2W.txt};
\addlegendentry{$L = 4$, $W = 0.2$};

\addplot[black, thick, dashed, mark=o, mark options={solid}, mark size=1.5pt]
table [x=mNak, y=Pe_mqam_fas, col sep=comma] {Figures/files_txt/plot7/sep_m_mqam_fas_16M_10gamdb_1Npe_4L_3Lhat_0.8W.txt};
\addlegendentry{$L = 4$, $W = 0.8$};

\addplot[black, thick, densely dotted, mark=square, mark options={solid}, mark size=1.5pt]
table [x=mNak, y=Pe_mqam_fas, col sep=comma] {Figures/files_txt/plot7/sep_m_mqam_fas_16M_10gamdb_1Npe_8L_3Lhat_0.2W.txt};
\addlegendentry{$L = 8$, $W = 0.2$};

\addplot[black, thick, dash dot, mark=triangle, mark options={solid}, mark size=1.5pt]
table [x=mNak, y=Pe_mqam_fas, col sep=comma] {Figures/files_txt/plot7/sep_m_mqam_fas_16M_10gamdb_1Npe_8L_3Lhat_0.8W.txt};
\addlegendentry{$L = 8$, $W = 0.8$};
\addplot[black, thick, mark=asterisk, mark options={solid}, mark size=1.5pt]
table [x=mNak, y=Pe_mqam_fas, col sep=comma] {Figures/files_txt/plot7/sep_m_mqam_fas_16M_10gamdb_10Npe_4L_3Lhat_0.2W.txt};

\addplot[black, thick, dashed, mark=o, mark options={solid}, mark size=1.5pt]
table [x=mNak, y=Pe_mqam_fas, col sep=comma] {Figures/files_txt/plot7/sep_m_mqam_fas_16M_10gamdb_10Npe_4L_3Lhat_0.8W.txt};

\addplot[black, thick, densely dotted, mark=square, mark options={solid}, mark size=1.5pt]
table [x=mNak, y=Pe_mqam_fas, col sep=comma] {Figures/files_txt/plot7/sep_m_mqam_fas_16M_10gamdb_10Npe_8L_3Lhat_0.2W.txt};

\addplot[black, thick, dash dot, mark=triangle, mark options={solid}, mark size=1.5pt]
table [x=mNak, y=Pe_mqam_fas, col sep=comma] {Figures/files_txt/plot7/sep_m_mqam_fas_16M_10gamdb_10Npe_8L_3Lhat_0.8W.txt};
\node[draw, semithick, ellipse, minimum width=0.2cm, minimum height=0.8cm, inner sep=0pt] at (axis cs:9,4e-4) {};
\node[black, font=\scriptsize, anchor=center] at (axis cs:9,5e-2) {$N_{\text{p}} \varepsilon = 1$};
\node[draw, semithick, ellipse, minimum width=2cm, minimum height=0.2cm, inner sep=0pt, rotate=45] at (axis cs:6.8,8e-9) {};
\node[black,font=\scriptsize, anchor=center] at (axis cs:5.75,5e-12) {$N_{\text{p}} \varepsilon = 10$};
\end{groupplot}
\end{tikzpicture}\vspace{-5mm}
\caption{Average SEP with $16$-PSK and $16$-QAM of a GSC$(3,L)$-based FAS receiver versus $m$-parameter, with $\overline{\gamma}=10$ dB and different values of $L$, $N_{\text{p}}\varepsilon$, and $W$.}
\vspace{-3mm}
\label{f3}
\end{figure}
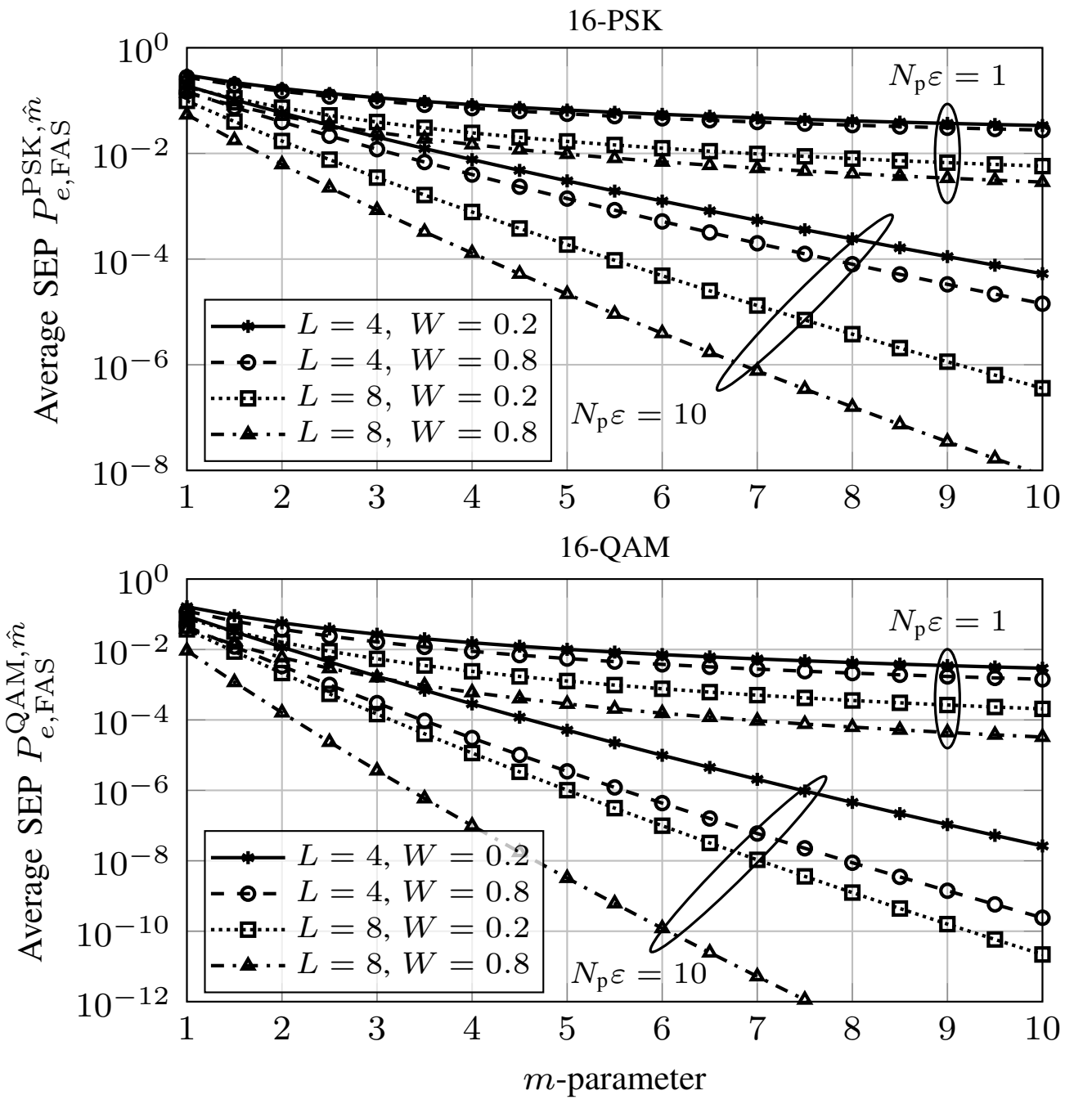

\section{Numerical Results and Discussion}
\label{Sec:Numerical}
This section presents a numerical evaluation of the derived analytical SEP expressions for the considered GSC-based FAS receiver over Nakagami-$m$ fading channels following Jakes' correlation model in \eqref{eq:CM_FAS}, with practical pilot-aided CE. We consider $M$-PSK and $M$-QAM signaling with $M=16$ and equal fading parameters for all $L$ FAS ports (or antenna elements for the case of MRC), i.e., $m_{\ell}=m$, $\forall \ell$. The analytical expressions are validated against Monte Carlo simulations averaged over $10^5$ independent channel realizations.

\begin{figure}[!t]
\centering
\pgfdeclarelayer{background}
\pgfdeclarelayer{foreground}
\pgfsetlayers{background,main,foreground}

\begin{tikzpicture}

\begin{groupplot}[
    group style={
        group size=1 by 1,
        vertical sep=25pt,
        y descriptions at=edge left,
    },
    width=8.5cm,
    height=5.0cm,
    xlabel near ticks,
    ylabel near ticks,
    ylabel style={yshift=-3pt},
    label style={font=\footnotesize},
    ticklabel style={font=\footnotesize},
    legend style={
        at={(0.02,0.02)},
        anchor=south west,
        font=\scriptsize,
        inner sep=1pt,
        row sep=-2pt,
        legend columns=2,
        fill opacity=0.75,
        draw opacity=1,
        text opacity=1
    },
    legend cell align=left,
    grid=major,
]
\nextgroupplot[
    xlabel={CE resources $N_{\text{p}} \varepsilon$},
    ylabel={$P_{e,{\rm FAS}}^{{\rm PSK},\hat{m}}$ and $P_{e,{\rm FAS}}^{{\rm QAM},\hat{m}}$},
    xmin=1, xmax=1e5,
    xtick={1,1e1,1e2,1e3,1e4,1e5}, xmode=log,
    ymin=1e-5, ymax=1, ymode=log,
    ytick={1,1e-2,1e-4,1e-6},
    title style={font=\scriptsize,yshift=-2mm},
]



\addplot[black, thick, mark=asterisk, mark options={solid}, mark size=1.5pt]
table [x=Npe, y=Pe_mpsk_fas, col sep=comma] {Figures/files_txt/plot9/sep_npe_mpsk_fas_16M_10gamdb_1m_8L_2Lhat_0.2W.txt};
\addlegendentry{$m = 1$, $16$-PSK};

\addplot[black, thick, dash dot, mark=o, mark options={solid}, mark size=1.5pt]
table [x=Npe, y=Pe_mpsk_fas, col sep=comma] {Figures/files_txt/plot9/sep_npe_mpsk_fas_16M_10gamdb_2.5m_8L_2Lhat_0.2W.txt};
\addlegendentry{$m = 2.5$, $16$-PSK};

\addplot[black, thick, dashed, mark=square, mark options={solid}, mark size=1.5pt]
table [x=Npe, y=Pe_mqam_fas, col sep=comma] {Figures/files_txt/plot9/sep_npe_mqam_fas_16M_10gamdb_1m_8L_2Lhat_0.2W.txt};
\addlegendentry{$m = 1$, $16$-QAM};

\addplot[black, thick, densely dotted, mark=triangle, mark options={solid}, mark size=1.5pt]
table [x=Npe, y=Pe_mqam_fas, col sep=comma] {Figures/files_txt/plot9/sep_npe_mqam_fas_16M_10gamdb_2.5m_8L_2Lhat_0.2W.txt};
\addlegendentry{$m = 2.5$, $16$-QAM};
\end{groupplot}
\end{tikzpicture}\vspace{-2mm}
\caption{Average SEP with $16$-PSK and $16$-QAM of a GSC$(2,8)$-based FAS receiver versus CE resources, with $\overline{\gamma}=10$ dB, $W=0.2$, and different values of $m$.}
\vspace{-3mm}
\label{f4}
\end{figure}
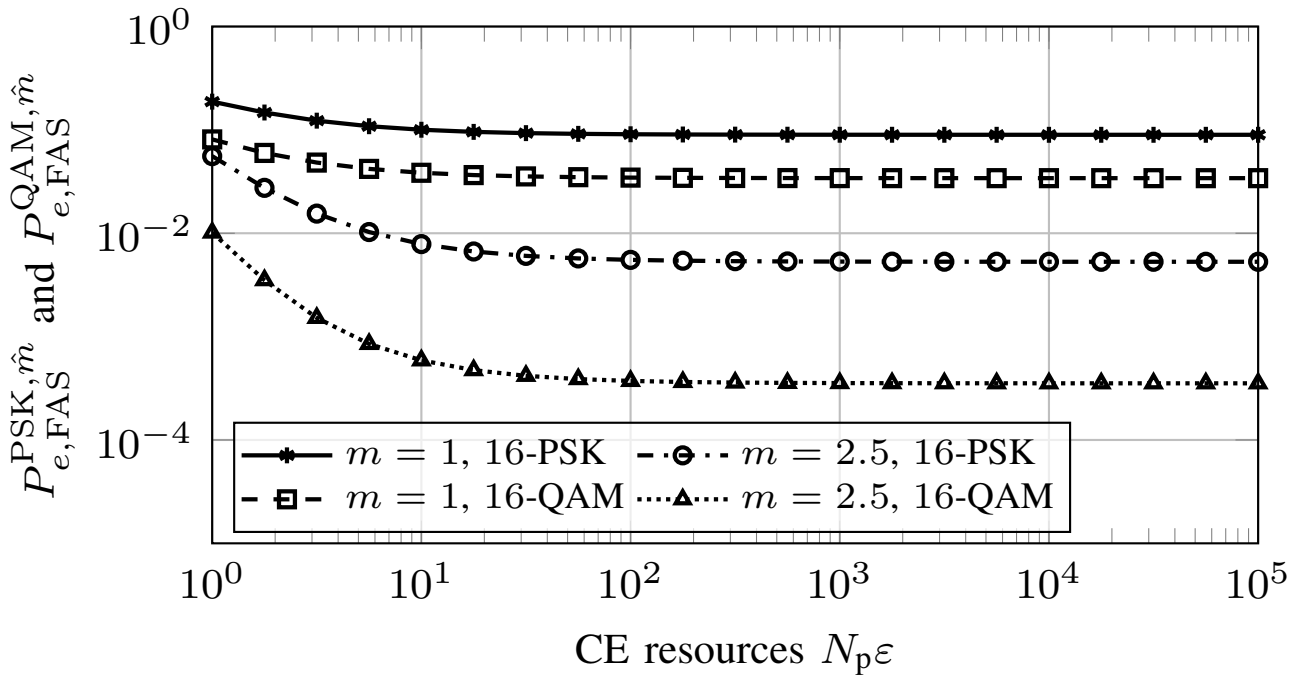

\begin{figure}[!t]
\centering
\input{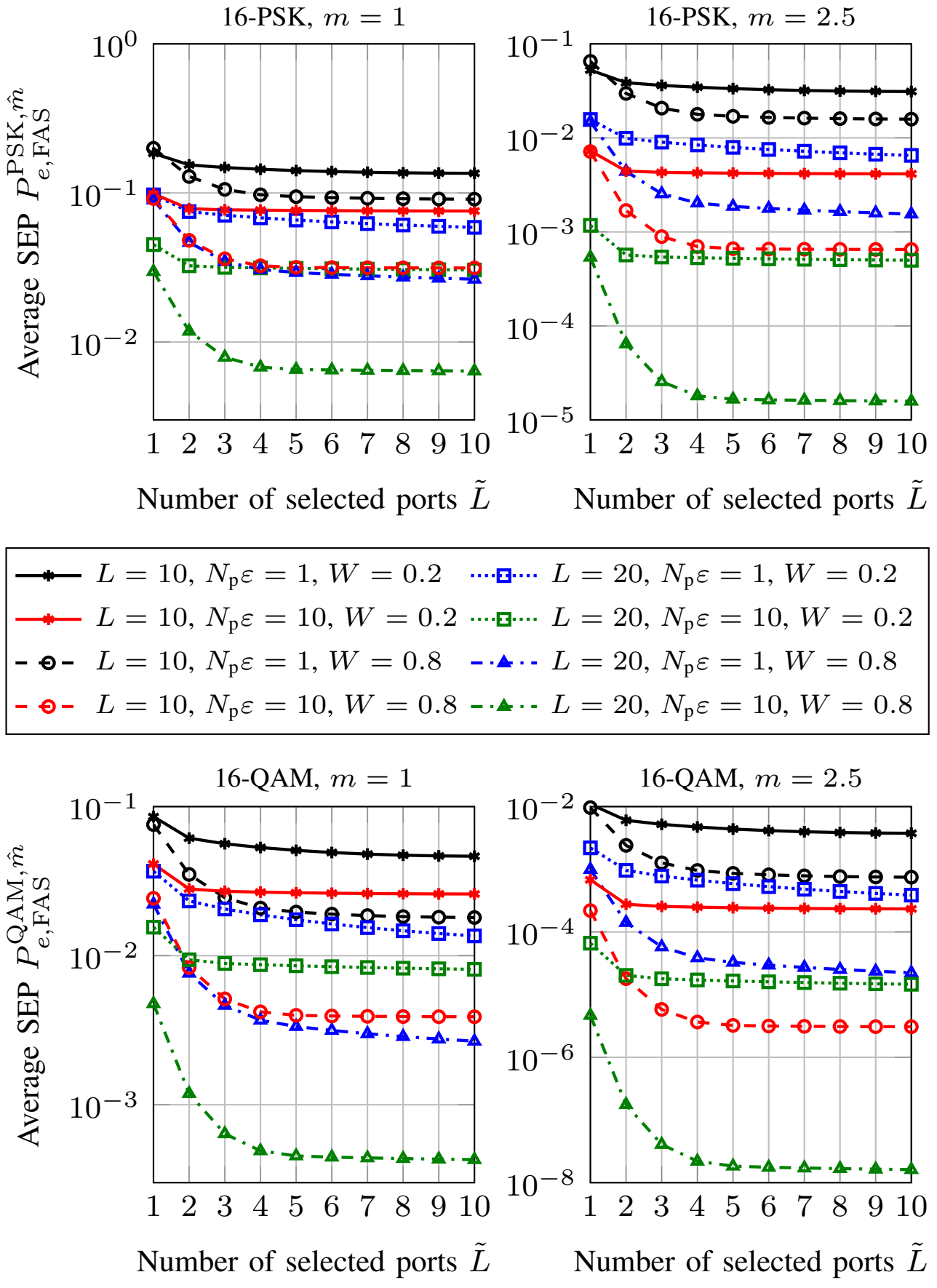} 
\caption{Average SEP with $16$-PSK and $16$-QAM of a GSC$(\tilde{L},L)$-based FAS receiver versus number of selected ports, with $\overline{\gamma}=10$ dB and different values of $m$, $L$, $N_{\text{p}}\varepsilon$, and $W$.}
\vspace{-5mm}
\label{f5}
\end{figure}

Fig.~\ref{f2} illustrates the average SEP of a GSC$(1,L)$-based FAS receiver versus the average SNR per symbol $\overline{\gamma}$, considering $m \in \{1, 2.5 \}$, $L \in \{2,8 \}$, $N_{\text{p}} \varepsilon \in \{1,10 \}$, and $W=0.2$. Specifically, the plot includes the expressions in \eqref{eq:sep_mpsk_fas_same_m_final}, \eqref{eq:sep_mqam_fas_same_m_final}, \eqref{eq:sep_fas_mpsk_high_snr}, \eqref{eq:sep_fas_mqam_high_snr}, and the case of perfect CSI (obtained by considering $N_{\text{p}} \varepsilon \to \infty$). The analytical expressions are in close agreement with their respective Monte Carlo simulations. Moreover, the presence of large CE errors and fading correlation (evident with increasing $L$ for fixed $W$) severely degrades the SEP over the whole range of $\overline{\gamma}$. In particular, the impact of CE becomes more severe as $m$ decreases and/or $L$ increases. Furthermore, consistent with the derived asymptotic analysis, all curves for a given $L$ converge to the same slope at high average SNR, confirming that the diversity order, governed solely by $L$ and $\hat m$, is preserved regardless of CE quality or fading correlation, while the constant vertical offset between curves reflects the coding-gain loss predicted by \eqref{eq:sep_fas_mpsk_high_snr} and \eqref{eq:sep_fas_mqam_high_snr}. The asymptotic behavior at large $L$ is only achieved at very high $\overline{\gamma}$ and is thus omitted in this figure.

Fig.~\ref{f3} plots the average SEP of a GSC$(3,L)$-based FAS receiver versus the IoHI $m$-parameter, considering $\bar\gamma = 10$~dB, $L \in \{ 4,8\}$, $N_{\text{p}}\varepsilon \in \{ 1,10\}$, and $W \in \{0.2, 0.8 \}$. In agreement with the diversity-order scaling $\hat{m} L$ established analytically, increasing $m$ yields a substantially steeper SEP reduction than a comparable increase in $L$. Specifically, moving from $L=4$ to $L=8$ at fixed $W$ produces a markedly smaller improvement than moving from $m=1$ to $m=2.5$ at fixed $L$, particularly for $16$-QAM due to its higher error-floor sensitivity. A larger $W$ (weaker Jakes' correlation across ports) consistently improves performance at every $m$, with the gap most pronounced at large $L$. On the other hand, a smaller $W$ incurs a persistent penalty even for large $L$, reflecting the fundamentally limited number of resolvable spatial degrees of freedom available in a densely packed FAS.

\begin{figure}[!t]  
\centering
\input{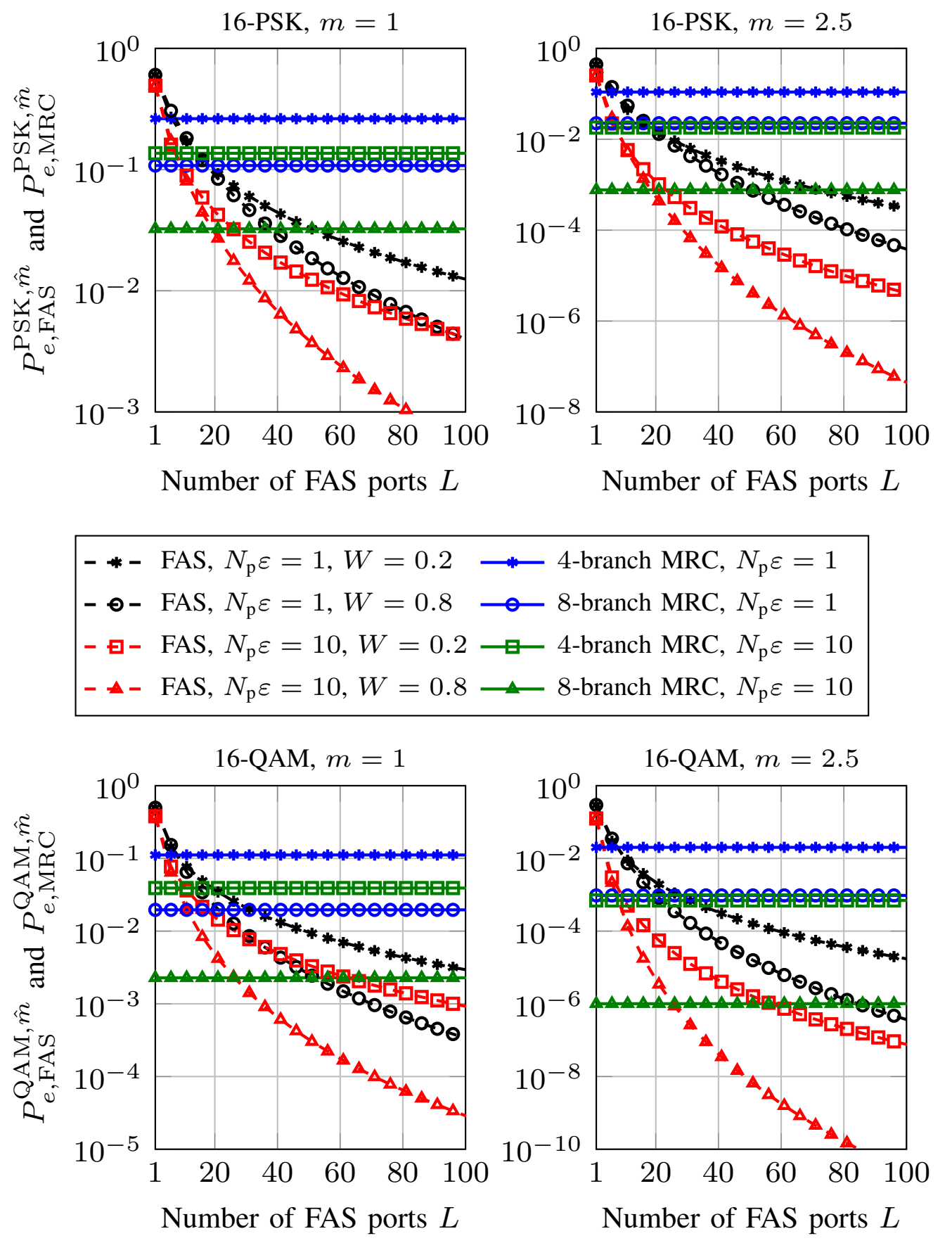}
\caption{Comparison of the average SEP with $16$-PSK and $16$-QAM between a GSC$(1,L)$-based FAS receiver and conventional $4$- and $8$-branch MRC receivers versus number of FAS ports, with $\overline{\gamma}=10$ and different values of $m$ and $N_{\text{p}}\varepsilon$.}
\vspace{-5mm}
\label{f6}
\end{figure}

\begin{figure*}[!t]
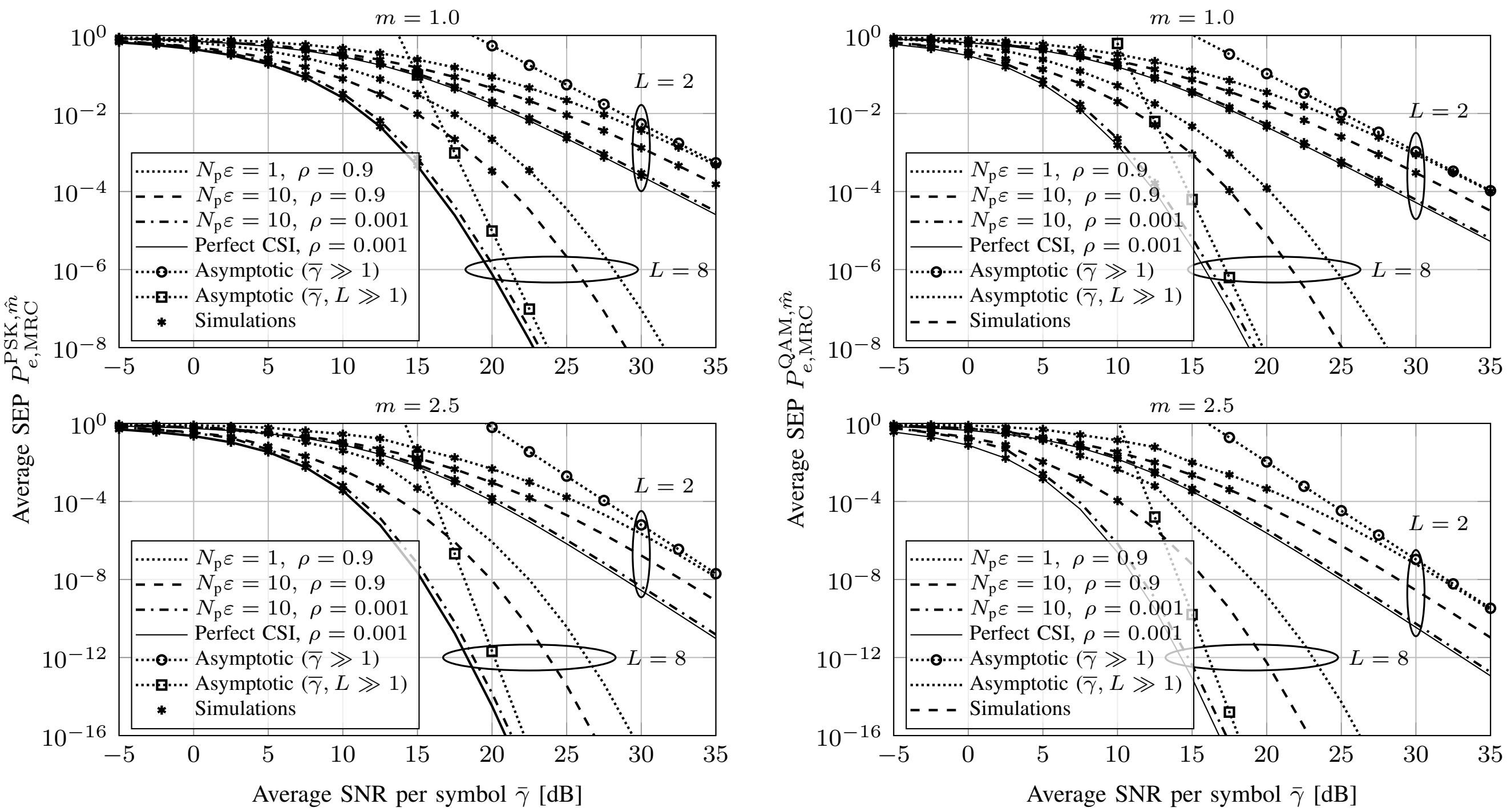

\begin{subfigure}[b]{\columnwidth}
\centering
\input{Figures/plot2.tex}\vspace{-2mm}
\caption{$16$-PSK.}
\label{f7a}
\end{subfigure}
\begin{subfigure}[b]{\columnwidth}
\centering
\input{Figures/plot3.tex}\vspace{-2mm}
\caption{$16$-QAM.}
\label{f7b}
\end{subfigure}
\caption{Average SEP with (a) $16$-PSK and (b) $16$-QAM of an $L$-branch MRC receiver versus average SNR per symbol, with different values of $m$, $L$, $N_{\text{p}}\varepsilon$, and $\rho$.}
\vspace{-3mm}
\label{f7}
\end{figure*}

Fig.~\ref{f4} illustrates the average SEP of a GSC$(2,8)$-based FAS receiver versus the CE resources $N_{\text{p}}\varepsilon$, considering $\overline{\gamma} = 10$~dB, $m \in \{ 1, 2.5 \}$, and $W=0.2$. All curves decrease monotonically and saturate toward the perfect-CSI floor as $N_{\text{p}}\varepsilon \to \infty$, consistent with $\eta_{\rm PSK} \to c_{\rm PSK} \overline{\gamma}$ and $\eta_{i,\rm QAM} \to c_{\rm QAM} \overline{\gamma}$, $\forall i$, in this limit. The curves with $m=2.5$ saturate at a lower $N_{\text{p}}\varepsilon$ compared with their counterparts with $m=1$, indicating that channels with lower fading severity are comparatively more sensitive to residual CE error and thus benefit more from additional pilot energy. This equivalently implies that achieving a target SEP under Rayleigh fading requires proportionally fewer pilot resources to approach the perfect-CSI floor than under milder (higher-$m$) fading. Furthermore, $16$-QAM exhibits a consistently steeper descent than $16$-PSK across the considered range, reflecting its higher sensitivity to the mismatched-detection penalty (i.e., the SNR degradation caused by detecting with a noisy weight instead of the true channel) captured by $\{\eta_{i,\rm QAM}\}$ relative to $\eta_{\rm PSK}$.

Fig.~\ref{f5} plots the average SEP of a GSC$(\tilde{L},L)$-based FAS receiver versus the number of selected ports $\tilde{L}$, considering $\overline{\gamma}=10$~dB, $m \in \{1,2.5 \}$, $L \in \{10,20 \}$, $N_{\text{p}} \varepsilon \in \{1,10 \}$, and $W \in \{0.2,0.8 \}$. All the SEP curves exhibit a monotonically decreasing behavior with $\tilde L$, but saturate at higher values of $\tilde{L}$. Specifically, moving from $\tilde{L}=1$ to $\tilde{L}=2$ produces a substantially larger improvement than any subsequent increment, since additional selected ports beyond the first few contribute increasingly correlated (and hence redundant) signal energy under Jakes' model. The gap between the curves corresponding to $W=0.2$ and $W=0.8$ narrows as $\tilde{L}$ grows, indicating that a wider length allows the receiver to extract most of the available diversity with fewer selected ports, whereas a compact aperture requires combining a larger fraction of the available $L$ ports to achieve comparable performance. This directly exposes the inherent trade-off between hardware/RF-chain complexity (governed by $\tilde{L}$) and the antenna's physical footprint (governed by $W$).

Fig.~\ref{f6} provides a comparison of the average SEP between a GSC$(1,L)$-based FAS receiver and conventional $4$- and $8$-branch MRC receivers (denoted by $P_{e, {\rm MRC}}^{{\rm PSK},\hat{m}}$ and $P_{e, {\rm MRC}}^{{\rm QAM},\hat{m}}$ in the figures) versus the number of FAS ports $L$, considering $\bar\gamma = 10$~dB, $m \in \{1, 2.5 \}$, and $N_{\text{p}} \varepsilon \in \{1,10 \}$. For smaller values of $L$, the FAS performs worse than both MRC benchmarks, since a single selected port cannot exploit multi-branch combining gain, and Jakes' correlation further limits the effective diversity among the $L$ candidate ports. However, as $L$ increases, the FAS curves improve steadily and eventually cross the MRC benchmarks. The crossing point occurs at a substantially larger $L$ for $W=0.2$ than for $W=0.8$, quantifying how many densely packed ports are required to match the performance of a small number of independent, fully-combined antenna branches. Thus, the number of ports needed to match a given MRC configuration is governed jointly by the length $W$ and the $m$-parameter, with the crossing point shifting to smaller $L$ as $m$ increases.

As described in Section \ref{Sec:Statistics}, the derived joint statistics are valid for any arbitrary power CM $\boldsymbol{\Sigma}$. They therefore also enable an analysis of the joint effect of CE errors and correlation among the fading channels on conventional MRC reception, which has also not been addressed in the existing literature. To this end, Fig.~\ref{f7} illustrates the average SEP of an $L$-branch MRC receiver versus the average SNR per symbol $\overline{\gamma}$, considering $m \in \{1, 2.5 \}$ and $N_{\text{p}} \varepsilon \in \{1,10 \}$. We adopt the exponential correlation model $\sqrt{\varrho_{k,\ell}}=\rho^{|k-\ell|}$~\cite{exp_corr2010} with $\rho\in[0,1)$ for the CMs of the actual Nakagami-$m$ faded channels. The plots are generated using the eigenvalues of the corresponding CM and by considering $\tilde{L}=L$ in the derived results. Similar to Fig.~\ref{f2}, the presence of both high CE errors and fading correlation severely degrades the SEP for all considered values of $\overline{\gamma}$. Moreover, the impact of CE becomes more severe as $\rho$ and/or $m$ decrease and/or $L$ increases. Furthermore, all curves for a given $L$ converge to the same slope at high average SNR, confirming that the diversity order, governed solely by $L$ and $\hat m$, is preserved regardless of the CE quality or fading correlation for conventional MRC reception as well. Lastly, the asymptotic curves at large $L$ become tight to the exact expressions beyond roughly $15$--$20$~dB, and the large-$L$ approximations remain accurate even for the modest number of branches considered here.
\section{Conclusions} \label{Sec:Conclusion}
In this paper, the joint statistics of estimated Nakagami-$m$ fading channels with arbitrary correlation were derived and exploited to analyze the SEP performance of a GSC-based FAS receiver that selects a subset of its available ports to combine them using MRC for data detection. A novel tight moment-matching-based Gamma approximation was obtained for the marginal distribution of the ML-estimated channel gains, together with a closed-form expression for the correlation coefficient between any pair of the estimated channel gains. Building on this framework, exact (and closed-form for identically distributed channels) SEP expressions were derived for a GSC-based FAS receiver under both $M$-PSK and $M$-QAM signaling, which were validated through Monte Carlo simulations across a wide range of fading, correlation, and CE-quality conditions. The derived high-SNR asymptotic expressions further established that non-ideal CE and fading correlation degrade only the achievable coding gain, while the diversity order remains intact and is dictated solely by the number of selected ports and the $m$-parameter. Numerical results demonstrated that the performance penalty due to imperfect CSI becomes increasingly severe as the correlation and/or $m$-parameter decrease and/or the number of diversity branches increases. It was also shown that the considered FAS receiver requires a sufficiently large number of densely packed ports and a sufficiently large aperture to match the performance of a conventional MRC receiver under identical CE conditions.
\bibliographystyle{IEEEtran}\vspace{-2mm}
\bibliography{references}
\balance
\end{document}